\documentclass[12pt]{article}
\usepackage{hyperref}
\usepackage[english]{babel}
\usepackage{amsmath}
\usepackage{latexsym}
\usepackage{amssymb}
\usepackage[all]{xy}
\usepackage[dvips]{graphicx}
\usepackage{epsfig}
\usepackage{psfrag}

\numberwithin{equation}{section} \numberwithin{table}{section}
\numberwithin{figure}{section}

\begin{document}



\begin{titlepage}

  \begin{center}

    \vspace{20mm}

    {\LARGE \bf A Holographic Model of Strange Metals}

    \vspace{10mm}

   Bum-Hoon Lee$^{\ast\dag}$, Da-Wei Pang$^{\sharp\ddag}$ and Chanyong Park$^{\dag}$

    \vspace{5mm}
    {\small \sl $\ast$ Department of Physics, Sogang University}\\
    {\small \sl Seoul 121-742, Korea\\}
    {\small \sl $\dag$ Center for Quantum Spacetime, Sogang University}\\
    {\small \sl Seoul 121-742, Korea\\}
    {\small \sl $\sharp$ CENTRA, Departamento de F\'{\i}sica,}\\
    {\small \sl Instituto Superior T\'{e}cnico,
Universidade T\'{e}cnica de Lisboa}\\
    {\small \sl Av. Rovisco Pais 1, 1049-001 Lisboa, Portugal}\\
    {\small \sl $\ddag$ Key Laboratory of Frontiers in Theoretical Physics,\\
    Institute of Theoretical Physics, Chinese Academy of Sciences, \\
    P.O. Box 2735, Beijing 100190, China\\}
    {\small \tt bhl@sogang.ac.kr, dwpang@gmail.com, cyong21@sogang.ac.kr}
    \vspace{10mm}

  \end{center}

\begin{abstract}
\baselineskip=18pt
We give a review on our recent work arXiv:1006.0779 [hep-th] and arXiv:1006.1719 [hep-th],
in which properties of holographic strange metals were investigated. The background is chosen to
be anisotropic scaling solution in Einstein-Maxwell-Dilaton theory with a Liouville potential.
The effects of bulk Maxwell field, an extra $U(1)$ gauge field and probe D-branes on
the DC conductivity, the DC Hall conductivity and the AC conductivity are extensively analyzed.
We classify behaviors of the conductivities according to the parameter ranges in the bulk theory
and characterize conditions when the holographic results can reproduce experimental data.
\end{abstract}
\setcounter{page}{0}
\end{titlepage}

\pagestyle{plain} \baselineskip=19pt

\tableofcontents

\section{Introduction}
The AdS/CFT correspondence~\cite{Maldacena:1997re, Aharony:1999ti}, or more generally
gauge/gravity duality, sets up a holographic duality between a weakly-coupled theory of
gravity in certain spacetime and a strongly-coupled field theory living on the boundary
of the spacetime. Thus gauge/gravity duality provides a powerful new
tool for studying strongly-coupled, scale-invariant field theories at finite charge
density. Hence one expects that this paradigm may be useful in condensed matter physics,
such as in understanding systems near quantum criticality~\cite{Hartnoll:2009sz}.
Actually the correspondence between gravity theories and condensed matter physics
(sometimes is also named as AdS/CMT correspondence)
has shed light on studying physics in the real world in the context of holography.

The simplest laboratory for studying the AdS/CMT correspondence
is Reissner-Nordstr\"{o}m-AdS(RN-AdS) black hole in four-dimensional spacetime.
The main reason why we consider such a background is that realistic condensed
matter systems always contain a finite charge density, which should be mapped into
charged AdS black holes in the gravity side according to gauge/gravity dictionary.
This laboratory has proven to be very efficient, for instance,
investigations of the fermionic two-point functions in this background indicate the existence of fermionic
quasi-particles with non-Fermi liquid behavior~\cite{Lee:2008xf, Liu:2009dm, Cubrovic:2009ye}, while the
$AdS_{2}$ symmetry of the extremal RN-AdS black hole is crucial to the emergent scaling symmetry
at zero temperature~\cite{Faulkner:2009wj}. Moreover, adding a charged scalar in such a background leads to superconductivity~\cite{Gubser:2008px, Hartnoll:2008vx, Hartnoll:2008kx}.

A further step towards a holographic model-building of strongly-coupled systems at
finite charge density is to consider the leading relevant (scalar) operator in the field theory side,
whose bulk gravity theory is an Einstein-Maxwell-Dilaton system with a scalar potential.
Such theories at zero charge density were analyzed in detail in recent years as they mimic
certain essential properties of QCD~\cite{Gursoy:2007cb, Gursoy:2007er, Gubser:2008yx, Gursoy:2008bu,
Gursoy:2008za}. Solutions at finite charge density have been considered in~\cite{Gubser:2009qt, Goldstein:2009cv, Gauntlett:2009bh, Cadoni:2009xm, Chen:2010kn, Goldstein:2010aw}
in the context of AdS/CMT correspondence.

Recently a general framework for the discussion of the holographic dynamics of
Einstein-Maxwell-Dilaton systems with a scalar potential was proposed in~\cite{Charmousis:2010zz},
which was a phenomenological approach based on the concept of Effective Holographic Theory (EHT).
The minimal set of bulk fields contains the metric $g_{\mu\nu}$, the gauge field $\mathcal{A}_{\mu}$
and the scalar $\phi$ (dual to the relevant operator). $\phi$ appears in two scalar functions that
enter the effective action: the scalar potential and the non-minimal Maxwell coupling. The main advantage
of this EHT approach is that it permits a parametrization of large classes of IR dynamics and allows
investigations on important observables. However, compared to models arising from string theory, the main
disadvantage is that it is not clear whether concrete EHTs have explicit string theory embeddings,
thus the corresponding field theories are less understood. Despite of its disadvantage,
the EHT is an effective method towards holographic model building of condensed matter systems. For subsequent generalizations see e.g.~\cite{Liu:2010ka, Lee:2010uy,
Gursoy:2010kw, Faulkner:2010gj, Bayntun:2010nx, AliAkbari:2010av, Astefanesei:2010dk,
Lee:2010ez, Pal:2010sx, Cadoni:2011kv, Meyer:2011xn, Gouteraux:2011xr}.

In this paper we will review our works on a holographic model of strange metals, which
were reported in~\cite{Lee:2010xx, Lee:2010ii}. Thermodynamic and
transport properties of strange metal phases, which have been observed in e.g.
heavy fermion compounds and high temperature superconductors, are of particular interest
in condensed matter physics. Some novel features of `strange metal' phases can be summarized
as follows:
\begin{itemize}
\item The DC resistivity is linear in temperature $T$ with $T$ much less than the
chemical potential $\mu$.
\item The AC conductivity behaves like $\sigma(\omega)\sim\omega^{-\nu}$ with $\nu\neq1$.
\item The Hall conductivity also exhibits an anomalous behavior $\sigma^{xx}/\sigma^{xy}\sim T^{2}$.
\end{itemize}
However, such puzzling behaviors are still less understood even at the theoretical level.

Recently a holographic model which can partly describe the above mentioned features of `strange metal'
phases was proposed in~\cite{Hartnoll:2009ns}. The background geometry was chosen to be
a four-dimensional asymptotic Lifshitz black hole,
\begin{equation*}
ds^{2}=L^{2}(-\frac{f(v)dt^{2}}{v^{2z}}+\frac{dv^{2}}{f(v)v^{2}}+\frac{dx^{2}+dy^{2}}{v^{2}}),
\end{equation*}
which possesses the following Lifshitz scaling symmetry
\begin{equation*}
t\rightarrow\lambda^{z}t,~~~v\rightarrow\lambda v,~~~\vec{x}\rightarrow\lambda\vec{x}
\end{equation*}
when $f(v)=1$, where $z$ is the dynamical exponent. In their settings the role of charge carriers was played by probe D-branes, whose dynamics can be described by the Dirac-Born-Infeld (DBI) action. The main results were
promising in some sense, as the DC resistivity and AC conductivity turned out to be
\begin{equation*}
\rho\sim T^{2/z},~~~\sigma(\omega)\sim\omega^{-2/z}, (z\geq2).
\end{equation*}
Comparing with experimental data
\begin{equation*}
\rho\sim T,~~~\sigma(\omega)\sim\omega^{-0.65},
\end{equation*}
it can be seen that the holographic calculations agree with the experimental data
at $z=2$ and $z=3$ respectively but not simultaneously. Furthermore, the Hall conductivity
took the following form
$$\sigma^{xx}/\sigma^{xy}\sim1/\sigma^{xx},$$
which mimic the Drude's theory rather than the strange metallic behavior.

However, as pointed out in~\cite{Hartnoll:2009ns}, considering a background with non-trivial
dilaton field may lead to more realistic model building for holographic strange metals. Actually
in~\cite{Charmousis:2010zz} the authors considered various exact solutions in Einstein-Maxwell-Dilaton
theory and characterized strange metallic behaviors by calculating the corresponding DC resistivity
and AC conductivity. In our papers~\cite{Lee:2010xx} and~\cite{Lee:2010ii}, we investigated
strange metallic behaviors in four-dimensional anisotropic black hole background, which is
an exact solution in Einstein-Maxwell-Dilaton theory with a Liouville type potential. In~\cite{Lee:2010xx}
we obtained the DC conductivity and AC conductivity by considering fluctuation of the $A_{x}$ component
of the bulk gauge field and classified the behaviors of the conductivity in different parameter regimes.
In~\cite{Lee:2010ii} we calculated conductivities induced by probe D-branes in such background and by
carefully choosing the parameters, we realized the DC resistivity and AC conductivity for strange metals
simultaneously. In addition, it should be pointed out that in~\cite{Kim:2010zq},
a quadratic temperature dependence of the inverse Hall angle in the presence of both electric and magnetic fields was
obtained by considering probe D7-branes in light-cone AdS black hole backgrounds.

The rest of the review is organized as follows: In section 2 we introduce the anisotropic solution
which is our playground. Then we calculate the DC and AC conductivities originating from fluctuations
of the bulk gauge field in section 3. The DC conductivity, AC conductivity and DC Hall conductivity
induced by probe D-branes are obtained in section 4. A summary will be presented in section 5.
\section{The background solution}

We start with the following action
\begin{equation}
S=\int d^{4}x\sqrt{-g}[R-2(\nabla\phi)^2-e^{2\alpha\phi}F_{\mu\nu}F^{\mu\nu}-V(\phi)],
\end{equation}
where $\phi$ and $V(\phi)$ represent the dilaton field and its potential.
Notice that the non-trivial dilaton coupling in front of the Maxwell term
may result in meaningful consequences. The equations of motion are given by
\begin{eqnarray}
R_{\mu\nu}-\frac{1}{2}Rg_{\mu\nu}+\frac{1}{2}g_{\mu\nu}V(\phi) &=& 2\partial_{\mu}\phi
\partial_{\nu}\phi-g_{\mu\nu}(\nabla\phi)^{2}+2e^{2\alpha\phi}F_{\mu\lambda}{F_{\nu}}^{\lambda}
-\frac{1}{2}g_{\mu\nu}e^{2\alpha\phi}F^{2}, \nonumber\\
\partial_{\mu}(\sqrt{-g}\partial^{\mu}\phi) &=& \frac{1}{4}\sqrt{-g}\frac{\partial
V(\phi)}{\partial\phi}+\frac{\alpha}{2}\sqrt{-g}e^{2\alpha\phi}F^{2},~~~\nonumber\\
0 &=& \partial_{\mu}(\sqrt{-g}e^{2\alpha\phi}F^{\mu\nu}).
\end{eqnarray}
Now, we fix the dilaton potential to be of the Liouville type
$V(\phi)=2\Lambda e^{-\eta\phi}$. When $\eta=0$, this potential
reduces to a cosmological constant and the corresponding solution was studied in
Ref. \cite{Goldstein:2009cv}.

To solve the equations of motion, we use
the following ansatz which leads to a zero temperature solution
\begin{equation}
ds^{2}=-a(r)^{2}dt^{2}+\frac{dr^{2}}{a(r)^{2}}+b(r)^{2}(dx^{2}+dy^{2}) ,
\end{equation}
with
\begin{equation}
a(r)=a_{0}r^{a_{1}},~~~b(r)=b_{0}r^{b_{1}},~~~\phi(r)=-k_{0}\log r.
\end{equation}
If we turn on the time component of the gauge field $A_{t}$ only,
the gauge field strength can be read off from the equation of motion
\begin{equation}
F_{tr}=\frac{q}{b(r)^{2}}e^{-2\alpha\phi}.
\end{equation}
Then we can fix the remaining parameters in the ansatz as follows by solving the equations of motion
\begin{eqnarray}
\label{2eq8}
 & &
a_{1}=1+\frac{k_{0}}{2}\eta,~~~b_{1}=\frac{(2\alpha-\eta)^{2}}{(2\alpha-\eta)^{2}+16},
~~~k_{0}=\frac{4(2\alpha-\eta)}{(2\alpha-\eta)^{2}+16}, ~~~b_{0}=1 ,\nonumber\\
& &
a_{0}^{2}=\frac{-2\Lambda}{(a_{1}+b_{1})(2a_{1}+2b_{1}-1)},~~~
q^{2}=- \left( \frac{2k_{0}}{a_{1}+b_{1}}+\frac{\eta}{2} \right)
\frac{\Lambda}{\alpha} ,
\end{eqnarray}
Notice that $b_1$ in~(\ref{2eq8}) is always smaller than $1$.

It can be seen that the above exact solution can reduce to various known solutions
in different limits. For example, when $\eta=0$ and $\Lambda =-3$ it
reduces to the one obtained in Ref. \cite{Goldstein:2009cv}. When
$2\alpha = \eta$, the above solution becomes
$AdS_2 \times R^2$. Furthermore, if we take the limit $\alpha \rightarrow \infty$ and at the
same time set $q=\eta=0$, we can obtain $AdS_4$ geometry. Finally, when
$\eta$ is proportional to $\alpha$, say $\eta = c \alpha$, the metric in the
limit $\alpha \to \infty$ reduces to a Lifshitz-like one~\cite{Kachru:2008yh, Taylor:2008tg}
\begin{equation}
ds^2 = - a_0^2 r^{2 z}
dt^2 + \frac{dr^2}{a_0^2 r^{2z}} + r^2 \left(dx^2 + dy^2  \right),
\end{equation}
with $z =(2+c)/(2-c)$. The exponent $z$ is given by $2$ for
$c=2/3$ and $3$ for $c=1$, etc.

The above zero temperature solution can be easily extended to a finite-temperature one
describing a black hole. With the same parameters in~(\ref{2eq8}), the black hole solution becomes
\begin{equation}
\label{2eq10}
ds^{2}=-a(r)^{2}f(r)dt^{2}+\frac{dr^{2}}{a(r)^{2}f(r)}+b(r)^{2}(dx^{2}+dy^{2}),
\end{equation}
with
\begin{equation}
f(r)=1-\frac{r^{2a_{1}+2b_{1}-1}_{+}}{r^{2a_{1}+2b_{1}-1}},
\end{equation}
where $r_+$ denotes the black hole horizon. Note that the $U(1)$ charge
is not explicitly shown in the metric components. The Hawking temperature of this black hole is given by
\begin{equation}
T \equiv  \frac{1}{4 \pi} \frac{\partial ( a(r)^2 f)}{\partial r}|_{r=r_+}
= \frac{(2 a_1 + 2 b_1 -1) a_0^2  r_+^{2 a_1 -1} }{4 \pi} ,
\end{equation}

\section{Strange metallic behavior from conventional approach}
As reviewed in the introduction, the DC and AC conductivities are observables which characterize
strange metallic behaviors. Thus it is desirable to work out the conductivities in the dual gravity background.
Generally, the conductivities can be evaluated in two complementary approaches. In this section we will
focus on the first approach, that is, we consider fluctuations of the $A_{x}$ component of the gauge field
and  calculate the conductivities via Kubo's formula, where the effective action for the fluctuations is described
by the conventional Maxwell term. The other approach, which involves $U(1)$ gauge fields on probe D-branes in
the dual background, will be considered in the next section.

For convenience, we introduce new coordinate variable
$u=r^{b_1}$. Then, the metric~(\ref{2eq10}) becomes
\begin{equation}
ds^2 = - g(u) f(u) e^{- \chi(u)} dt^2 + \frac{du^2}{g(u) f(u)} + u^2 (dx^2  + dy^2) ,
\end{equation}
where
\begin{equation}
g(u) =a_0^2 b_1^2 u^{2 (a_1 + b_1 -1)/b_1},~~~
e^{\chi(u)} = b_1^2 u^{2(b_1 -1)/b_1} ,
\end{equation}
and
\begin{equation}
f(u) = 1-\frac{u_+^{(2 a_1 + 2 b_1 -1 )/b_1} }{u^{(2 a_1 +2 b_1 -1)/b_1}} .
\end{equation}
\subsection{Turing on fluctuations of background gauge field}
It can be seen that once we turn on the gauge field fluctuation $A_x$, the metric fluctuations
$g_{tx}$ and $g_{ux}$ are also involved. However, the corresponding equations for the fluctuations
can be simplified in the gauge $g_{ux} = 0$ and $A_u=0$. Consider the Maxwell action
\begin{equation}
S = - \frac{1}{4} \int d^4 x \sqrt{- g} \ h^2  F_{\mu\nu}  F^{\mu\nu} ,
\end{equation}
where $h^2 = 4 e^{2 \alpha \phi}$ is a nontrivial coupling function and the following ansatz
\begin{equation}
A_x (t,u) = \int \frac{d\omega}{ 2 \pi} e^{- i\omega t } A_x (\omega,u),~~~
g_{tx}(t,u) = \int \frac{d\omega}{ 2 \pi} e^{- i\omega t }  g_{tx} (\omega,u) ,
\end{equation}
the equation of motion for $A_x$
\begin{equation}
0 = \frac{1}{\sqrt{-g}} \partial_{\mu}( \sqrt{-g} \ h^{2}
g^{\mu \rho} g^{x \nu} F_{\rho \nu})
\end{equation}
becomes
\begin{equation}
0 = \partial_u ( e^{- \frac{\chi}{2}} g f h^2 \partial_u A_x )
+ \omega^2 e^{\frac{\chi}{2}} \frac{h^2}{g f} A_x
+ e^{\frac{\chi}{2}} h^2 (\partial_u A_t) ( \partial_{u}g_{tx} - \frac{2}{u} g_{tx}).
\end{equation}
Moreover, the $(u,x)$-component of the Einstein equation is given by
\begin{equation}
\partial_{u}g_{tx}- \frac{2}{u} g_{tx} = - h^2 (\partial_u A_t) A_x .
\end{equation}

Combining the above two equations, we can obtain
\begin{equation}
\label{2eq15}
0 = \partial_u (e^{- \frac{\chi}{2}} g f h^2 \partial_u A_x)
+ \omega^2 e^{\frac{\chi}{2}} \frac{h^2}{g f} A_x
- e^{\chi/2} h^4 (\partial_u A_t)^2 A_x .
\end{equation}
Next we introduce the following new variables
\begin{equation}
\label{2eq16}
- \frac{\partial}{\partial v} = e^{- \frac{\chi}{2}} g \frac{\partial}{\partial u} ,~~~
A_x = \frac{\Psi}{ \sqrt{f} h},
\end{equation}
(\ref{2eq15}) simply reduces to a Schr\"{o}dinger-type equation
\begin{equation}
0 = \Psi'' + V(v) \Psi
\end{equation}
with the effective potential given by
\begin{equation}
\label{2eq17}
V(v) = (\omega^2 + \frac{(f')^2}{4}) \frac{1}{f^2}
- (\frac{f' h'}{h} + \frac{f''}{2} + \frac{h^2}{g} e^{\chi} (A_t')^2)\frac{1}{f}
- \frac{h''}{h} ,
\end{equation}
where the prime denotes derivative with respect to $v$.

One can express $v$ in terms of $u$ by simply reversing the first equation
in~(\ref{2eq16}). Here we will concentrate on the case $a_1 > 1/2$, in which
$v$ is given by
\begin{equation}
v = \frac{1}{(2 a_1 -1) \ a_0^2 \ u^{\frac{2 a_1 -1}{b_1}}} .
\end{equation}
Similarly the black hole horizon $v_+$ can be written as
\begin{equation}
v_+ = \frac{1}{(2 a_1 -1) \ a_0^2 \ u_+^{\frac{2 a_1 -1}{b_1}}} .
\end{equation}
Therefore in $v$-coordinate the boundary ($u=\infty$) is located at $v=0$ and
the zero temperature limit corresponds to setting the black hole horizon to $v=\infty$ ($u=0$).
\subsubsection{At zero temperature}
We first consider the zero temperature case, in which $f=1$ and $f'=f''=0$.
The effective potential~(\ref{2eq17}) reduces to
\begin{equation}
V(v) = \omega^2 -  \frac{h^2}{g} e^{\chi} (A_t')^2 - \frac{h''}{h} .
\end{equation}
Furthermore, the above effective potential
can be written in terms of the $v$ coordinate
\begin{equation}
\label{2eq18}
V(v) = \omega^2 - \frac{c}{v^2} ,
\end{equation}
with a constant $c$
\begin{equation}
c = \frac{4 (16 + 4 \alpha^2 - \eta^2)[8 + (2 \alpha- \eta) (\alpha
+ \eta)]}{ (16 + 4\alpha^2 + 4 \alpha \eta - 3 \eta^2)^2}.
\end{equation}

The exact solution $\Psi$ to the Schr\"{o}dinger equation can be expressed
in terms of the $i$-th kind of Hankel function $H^{(i)}$
\begin{equation}
\Psi = c_1 \sqrt{v} H^{(1)}_{\delta} (\omega v) + c_2 \sqrt{v} H^{(2)}_{\delta}(\omega v) ,
\end{equation}
where $c_1$ and $c_2$ are constants and
\begin{equation}
\label{2eq19}
\delta = \frac{\sqrt{1+ 4 c}}{2} .
\end{equation}
By imposing the incoming boundary condition at the horizon ($v=\infty$),
we can pick up the solution with $c_2 = 0$. Then the asymptotic expansion of the solution
near the boundary becomes
\begin{equation}
\Psi \approx \Psi_0 ( v^{\frac{1}{2} - \delta}
-  (\frac{\omega}{2})^{2 \delta} \frac{\Gamma (1 - \delta)}{\Gamma (1+\delta)} e^{- i \pi \delta} v^{\frac{1}{2}
+ \delta}).
\end{equation}
with
\begin{equation}
c_1 = \frac{i \pi }{\Gamma (\delta)} ( \frac{\omega}{2})^{\delta} \Psi_0 .
\end{equation}
Then we can obtain the solution for $A_x$ from~(\ref{2eq16}) and extract
the retarded Green's function $G^{R}_{xx}$ from the boundary action~\cite{Son:2002sd}. Finally
the AC conductivity is given by Kubo's formula
\begin{equation}
\sigma = \frac{G^{R}_{xx}}{i \omega} \sim  \omega^{2 \delta - 1} .
\end{equation}
For the DC conductivity we should set $\omega= 0$. Then the DC conductivity
of this system becomes infinity when $2 \delta < 1$ or zero when $2 \delta > 1$.
As reviewed in the introduction, if we choose $2 \delta -1 = - 0.65$.
the AC conductivity agrees with that of strange metals. Notice that since there are two free parameters,
$\alpha$ and $\eta$, when $\Lambda=-3$ we can find infinite many pairs of the parameters $(\alpha, \eta)$
which give $\sigma \sim\omega^{-0.65}$ and satisfy the regularity condition $q^2 > 0$, such as
$(\alpha,\eta) \approx (1,3.804), \ (2,5.196), \ (2,5.338), \cdots$.
\subsubsection{At finite temperature}
The calculations in the finite temperature background can be performed in a similar way.
At the horizon, the dominant term in the effective potential~(\ref{2eq17})is given by
\begin{equation}
U(v) \approx ( \omega^2 + \frac{(f_h')^2}{4}) \frac{1}{f_h^2}
= (\omega^2 +   \frac{(2 a_1 + 2 b_1 - 1)^2}{4 (2 a_1 -1)^2 v_+^2} ) \frac{1}{f_h^2} ,
\end{equation}
where $f_h$ denotes the value of $f$ at the horizon. Then we can obtain the approximate solution
\begin{equation}
\Psi = c_1 f^{\nu_-} + c_2 f^{\nu_+} ,
\end{equation}
where
\begin{equation}
\nu_{\pm} = \frac{1}{2}\pm i \sqrt{ \omega^2 + \frac{(2 a_1 + 2 b_1 - 1)^2}{4 (2 a_1 -1)^2 v_+^2}
- \frac{1}{4} } .
\end{equation}
Furthermore we can set $c_2 = 0$ by imposing the incoming boundary condition again.

\begin{figure}
\begin{center}
\vspace{-1cm}
\hspace{-0.5cm}
\includegraphics[angle=0,width=0.45\textwidth]{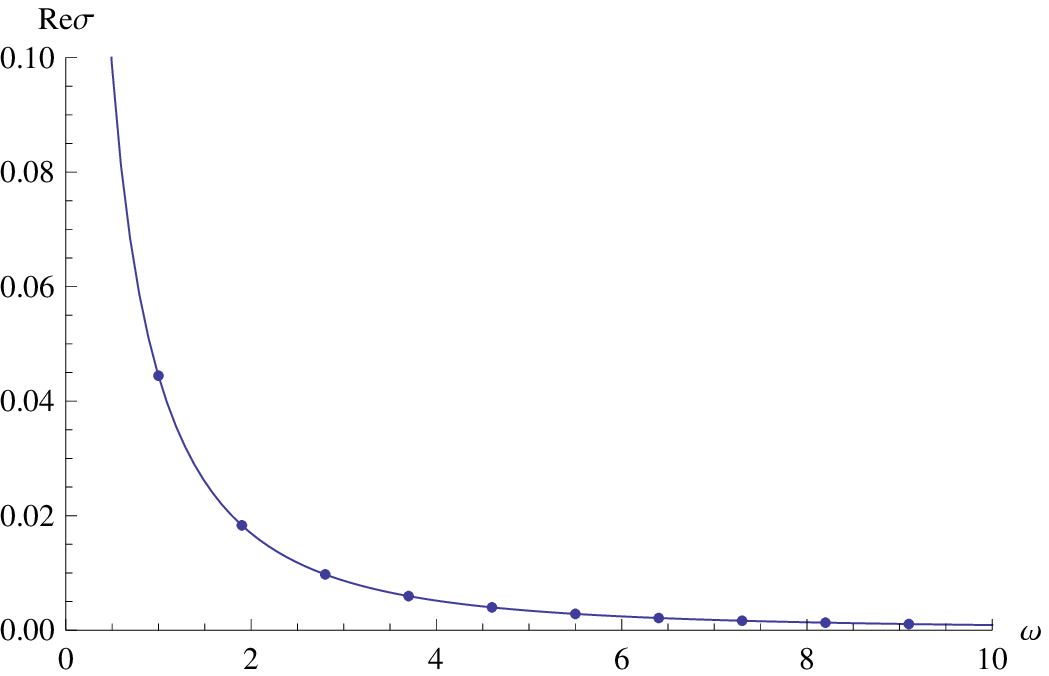}
\hspace{-0cm}
\includegraphics[angle=0,width=0.45\textwidth]{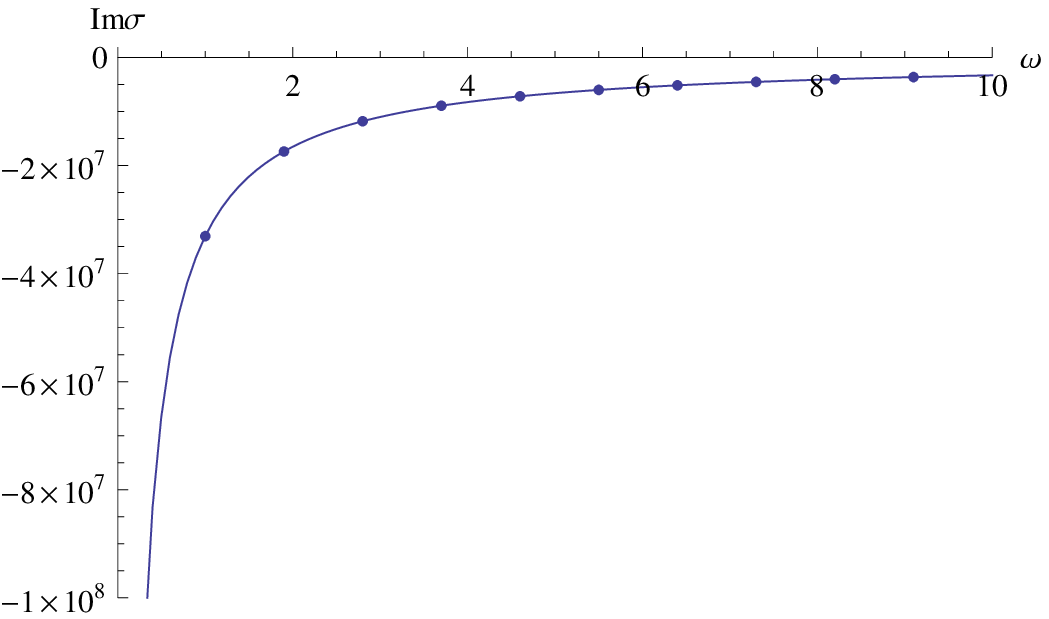}
\vspace{-0cm} \\
\caption{\small The conductivity at the finite temperature where we choose
$\alpha=2$, $\eta=1$ and $\Lambda=-3$.}
\label{pic1}
\end{center}
\end{figure}

Now, we investigate the asymptotic behavior of $\Psi$. Since the leading behavior of
the effective potential near the boundary is given by~(\ref{2eq18}), the perturbative
solution is
\begin{equation}
\Psi = d_1 \ v^{\frac{1}{2} - \delta} + d_2 \ v^{\frac{1}{2} + \delta},
\end{equation}
where $\delta$ has been defined in~(\ref{2eq19}).
The subsequent calculations are more or less the same as the zero temperature case, but it should
be pointed out that here we have made use of numerical techniques.
Finally the AC conductivity at finite temperature reads
\begin{equation}
\sigma = \frac{4 a_0^2}{i\omega} [ (2 a_1 -1) (\frac{1}{2} + \delta) + b_1   -  \alpha k_0]
\frac{ d_2 }{d_1 [ (2 a_1 - 1) a_0^2 ]^{2 \delta}} ,
\end{equation}
where the last factor $\frac{ d_2 }{d_1 [(2 a_1 - 1) a_0^2 ]^{2 \delta}}$ can be numerically
obtained by solving the Schr\"{o}dinger equation together with
the initial data at the horizon. The real and imaginary part of the AC conductivity are plotted
in Figure~\ref{pic1} .
We can fit the dual AC conductivity in Figure \ref{pic1}
with the following expected form
\begin{equation}
\label{2eq20}
\sigma = a \omega^{-b} .
\end{equation}
If we choose $\alpha=2$, $\eta=1$ and $\Lambda=-3$
for the simple numerical calculation, the conductivity
can be fitted by~(\ref{2eq20}) with
$a \approx 0.045$ and $b \approx 1.5$.
\subsection{The effect of an extra gauge field}
We will consider the effect of an extra $U(1)$ gauge field in this subsection.
Note that although the Maxwell coupling is trivial in this case,
we can find various different behaviors for the conductivities due to
the free parameters in the theory. For later convenience, we introduce a different radial coordinate
$z = 1/r$. Then the black hole metric can be rewritten as
\begin{equation}
ds^{2}=- \frac{a_0^2}{z^{2 a_1}} f(z) dt^{2}+\frac{ z^{2 a_1} dz^{2}}{a_0^2 z^4 f(z)}
+ \frac{dx^{2}+dy^{2}}{z^{2 b_1}}
\end{equation}
with
\begin{equation}
f(z) = 1 - \frac{z^{2 a_1 + 2 b_1 -1}} {z_+^{2 a_1 + 2 b_1 -1}},
\end{equation}
where all parameters are same as ones in~(\ref{2eq8}) and $z_+$ denotes
the location of the horizon. The Hawking temperature is given by
\begin{equation}
T = \frac{(2 a_1 + 2 b_1 -1) a_0^2 }{4 \pi z_+^{2 a_1 -1}} .
\end{equation}

Now, we introduce another $U(1)$ gauge field fluctuation $a_{\mu}$, which is not coupled with the dilaton field
\begin{equation}
S^{\prime} = - \frac{1}{4} \int d^4 x \sqrt{-g} f_{\mu\nu} f^{\mu\nu} ,
\end{equation}
where $f_{\mu\nu} = \partial_{\mu} a_{\nu} - \partial_{\nu} a_{\mu}$
and we have absorbed the gauge coupling constant to the gauge field.
After choosing $a_z = 0$ gauge and turning on the
$x$-component of the gauge fluctuation
\begin{equation}
a_x (x) = \int \frac{d\omega dk}{(2 \pi)^2} e^{- i\omega t + ikx} a_x (\omega,k,z),
\end{equation}
the corresponding Maxwell equation becomes
\begin{equation}
\label{eq55}
0 =  a_x'' + (\frac{2(1-a_1)}{z} + \frac{f'}{f}) a_x'
+ (\frac{\omega^2}{a_0^4 z^{4(1- a_1)} f^2}  -
\frac{k^2}{a_0^2 z^{4-2 a_1-2b_1} f}) a_x ,
\end{equation}
where the prime denotes derivative with respect to $z$.
\subsubsection{At zero temperature}
The calculations are essentially the same as those presented in previous subsection, so here we
just list the main results for each parameter regime.
\begin{itemize}
\item $1/2<a_{1}=b_{1}\leq1$\\
In this case, the equation of motion for $a_x$ becomes
\begin{equation}
 a_x'' + \frac{2 \delta}{z}  a_x' +  \frac{\gamma}{z^{4 \delta}}a_x=0,
\end{equation}
where
\begin{equation}
\delta= 1- a_1  \qquad {\rm and} \qquad \gamma = \frac{\omega^2}{a_0^4} - \frac{k^2}{a_0^2} ,
\end{equation}
with $0 \leq \delta < 1$. After imposing the incoming boundary condition,
the solution is given by
\begin{equation}
a_x = c_1  \exp ( i \frac{\sqrt{\gamma} }{1- 2 \delta} z^{1-2 \delta}) .
\end{equation}
For $\delta > 1/2$, we can not purturbatively expand this solution near the boundary ($z=0$). In
such a case it is unclear how to define the dual operator, so we consider only the case $\delta < 1/2$
(or $a_1 > 1/2 $) from now on. In this case, $a_x$ has the following expansion near the boundary
\begin{equation}
a_x = a_0 (1 + i \frac{\sqrt{\gamma} }{1- 2 \delta} z^{1-2 \delta} + \cdots ) ,
\end{equation}
where $a_0= c_1$ corresponds to the boundary value of $a_x$,
which can be identified with the source term
of the dual gauge operator.

\begin{figure}
\begin{center}
\vspace{1cm}
\hspace{-0.5cm}
\includegraphics[angle=0,width=0.5\textwidth]{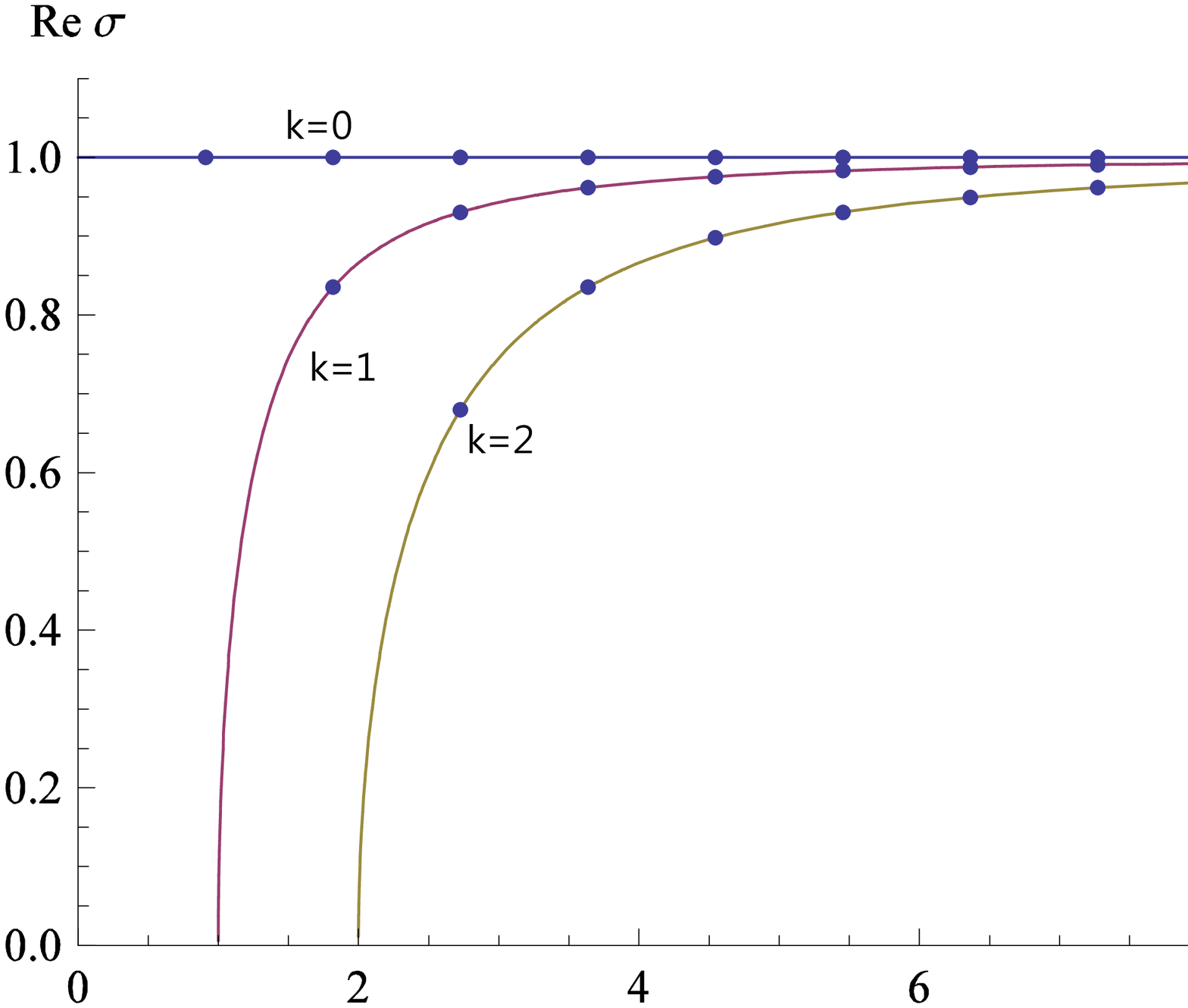}
\hspace{-0.5cm}
\includegraphics[angle=0,width=0.5\textwidth]{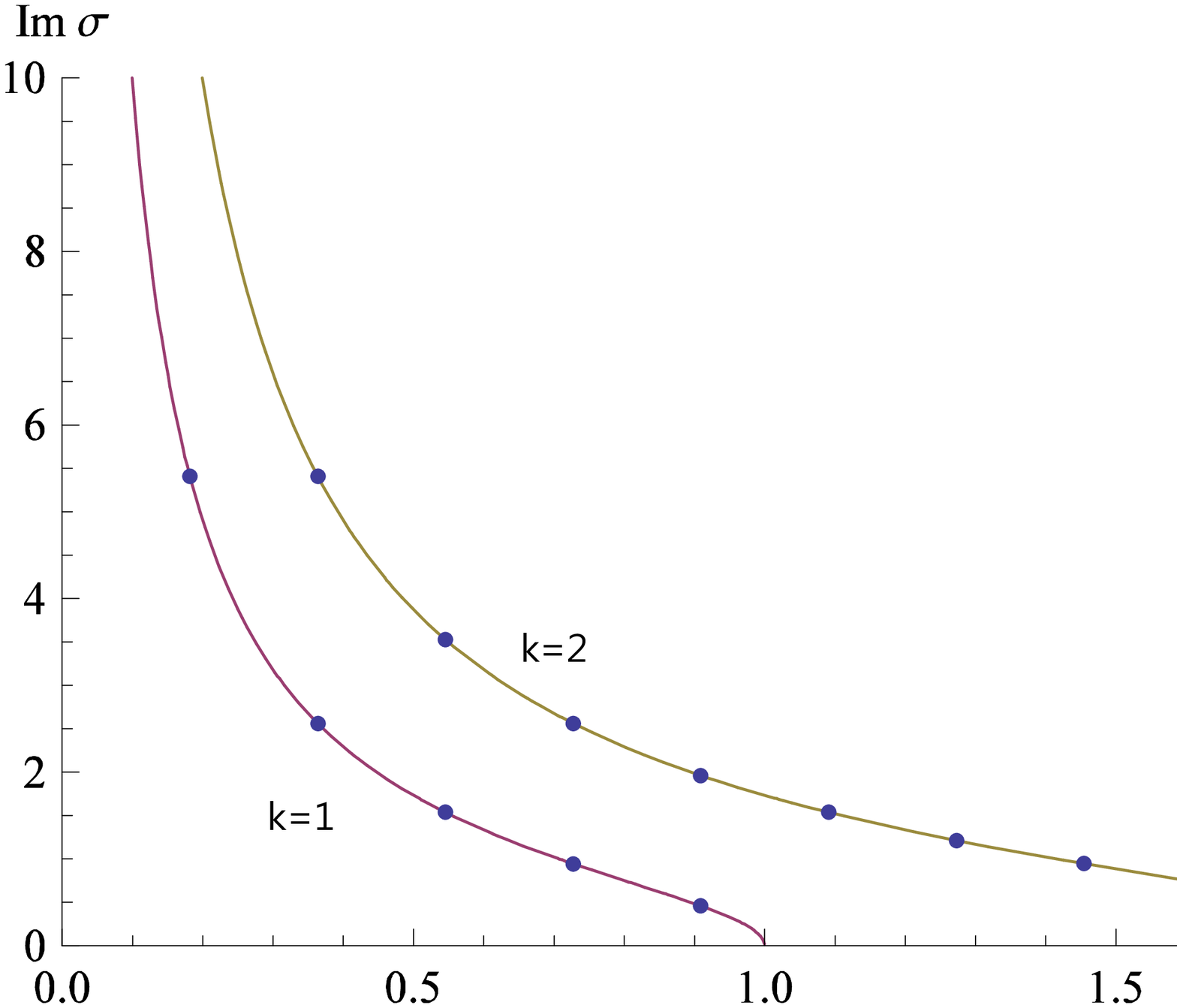}
\vspace{-2.5cm} \\
\caption{\small The real and imaginary conductivity at $a_0 = 1$ with $a_1 = b_1 \leq 1$.}
\label{fcawoc}
\end{center}
\end{figure}

After extracting the retarded Green's function, the conductivity is given by
\begin{equation}
\label{2eq21}
\sigma = \sqrt{1- \frac{a_0^2 k^2 }{\omega^2}} .
\end{equation}
For the time-like case ($ \omega^2 > k^2 a_0^2$), the conductivity is real.
In the space-like case, the imaginary conductivity appears.
In addition, the AC conductivity for $k=0$ becomes a constant $\sigma_{AC}  = 1$,
in which there is no imaginary part of the conductivity.
In Figure \ref{fcawoc}, we plot the real and imaginary conductivity, in which we can see that
as the momentum $k$ increases the real or imaginary conductivity decreases or increases
respectively. Furthermore, as shown in~(\ref{2eq21}) and figure 2,
the real and imaginary conductivities become zero at $\omega^2 = a_0^2 k^2$. Below or above
this critical point, there exists only the imaginary or real conductivity respectively.

\item $1/2<b_1<a_1\leq1$\\
It is impossible to solve the Maxwell equation analytically with arbitrary $a_1$ and $b_1$.
So we will try to obtain the retarded Green's function and conductivity numerically.
To do so, we should first find out the approximate
behavior of $a_x$ near the horizon as well as at the asymptotic boundary.
It can be seen that at the horizon, the approximate solution satisfying the
incoming boundary condition is given by
\begin{equation}
\label{2eq22}
a_x = c   \exp ( i \frac{\omega}{a_0^2 (2 a_1 -1)} z^{ 2 a_1 -1 }) .
\end{equation}
On the other hand, the two leading terms of the asymptotic solution near the boundary are
\begin{equation}
\label{eq66}
a_x = c_1 + c_2   z^{2 a_1 -1} ,
\end{equation}
where $c_1$ and $c_2$ are integration constants.

To find the relation between $c_1, c_2$ and $c$, we should solve the Maxwell equation numerically
with the initial conditions determined from~(\ref{2eq22}).
We omit intermediate steps and present the final result
for the conductivity
\begin{equation}
\sigma =  \frac{a_0^2 a_x' (\epsilon)}{ i \omega a_x (\epsilon) \epsilon^{2 a_1 - 2 }},
\end{equation}
where $\epsilon$($\epsilon \rightarrow 0$) denotes the UV cut-off.
In Figure \ref{figg3}, we plot the real and imaginary conductivity. Notice that
in this case there is no critical point like the $ 1/2 < a_1 = b_1 \leq 1$ case.
In other words, the real and imaginary conductivities are well defined in the whole
range of the frequency. In particular, for large $k$ the real conductivity becomes
zero as the frequency goes to zero. For $k=0$, the real conductivity is a constant
like the previous case. For $k \neq 0$ the conductivity grows as the frequency
increases, which is opposite to the strange metallic conductivity.

\begin{figure}
\begin{center}
\vspace{1cm}
\hspace{-0.5cm}
\includegraphics[angle=0,width=0.5\textwidth]{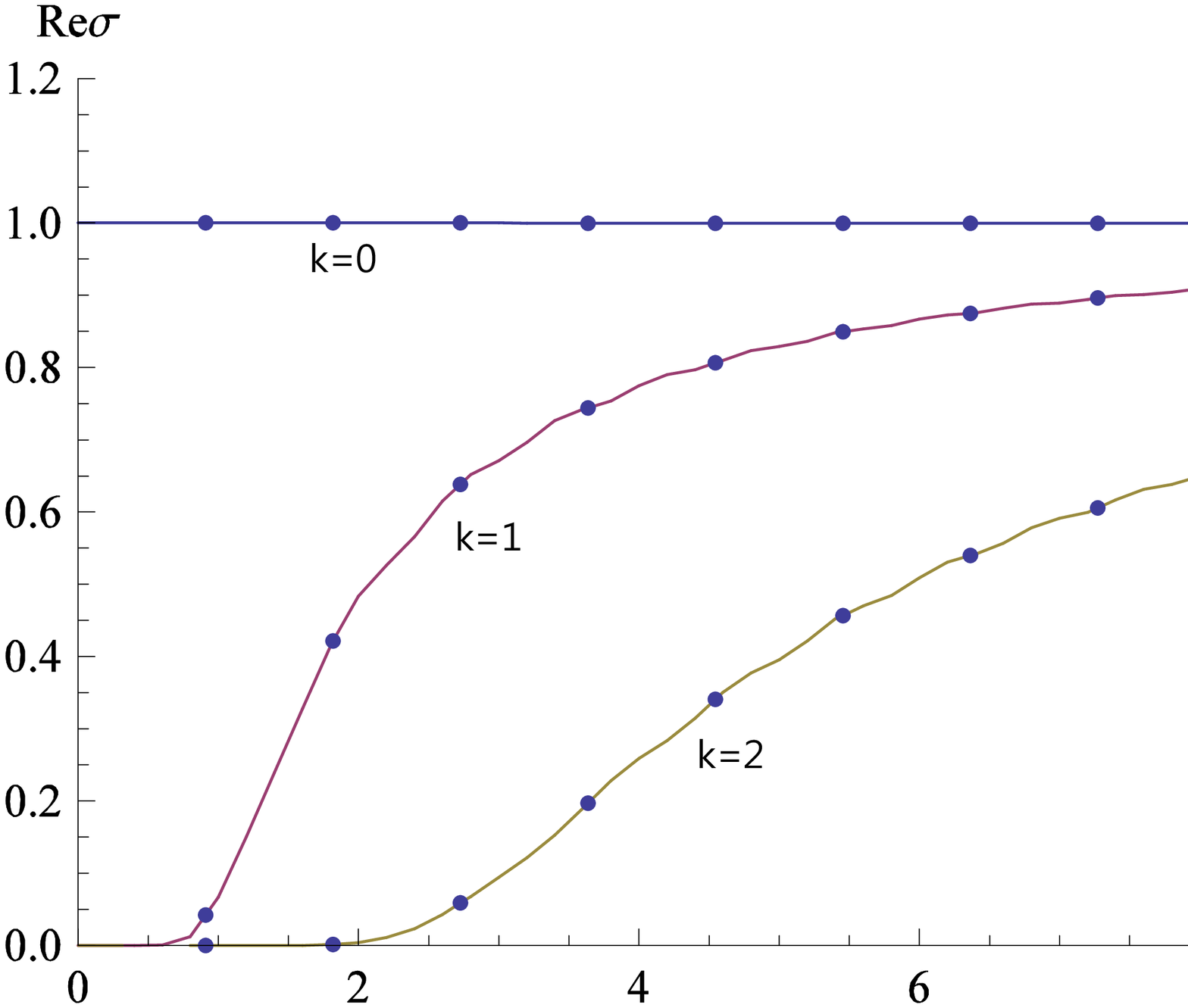}
\hspace{-0.5cm}
\includegraphics[angle=0,width=0.5\textwidth]{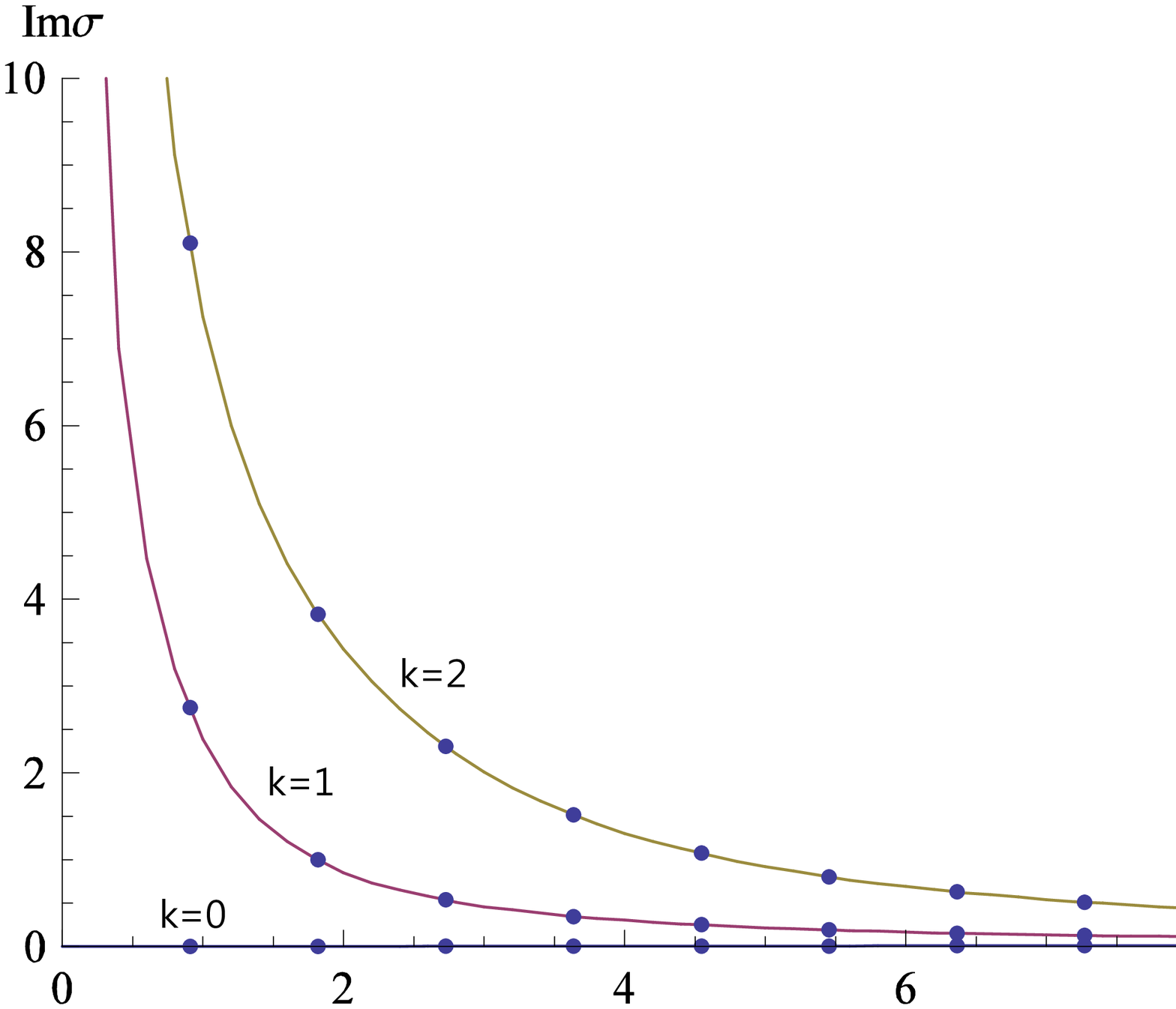}
\vspace{-2.5cm} \\
\caption{\small The real and imaginary conductivity at $a_0 = 1$ with $1/2 <b_1  < a_1 \leq 1$.}
\label{figg3}
\end{center}
\end{figure}
\item $1/2<a_1<b_1\leq1$\\
In this case, the $k^2$ term in~(\ref{eq55}) is dominant at the horizon. Since
the near horizon behavior of this solution is space-like, we should
impose the regularity condition instead of the incoming boundary condition.
The exact solution is given by
\begin{equation}
a_x =  z^{a_1 - \frac{1}{2}} (d_1' I_{- \nu} ( x )  + d_2' I_{\nu} ( x )),
\end{equation}
with two integration constants $d_1'$ and $d_2'$, where
$I_{\nu} (x)$ is the modified Bessel function and
\begin{equation}
\nu = \frac{2 a_1 -1}{2 (a_1 + b_1 -1)} \qquad {\rm and} \qquad
x = \frac{k z^{a_1 + b_1 -1}}{a_0 (a_1 + b_1 -1)} .
\end{equation}
After picking up the regular solution at the horizon, we can arrive at
the following expansion
\begin{equation}
a_x = \frac{d_1  \exp ( - \frac{k z^{a_1 + b_1-1}}{a_0 (a_1 + b_1 -1) } )}{z^{(b_1 - a_1)/2}},
\end{equation}
while the near boundary solution is still given by the same expression
~(\ref{eq66}).

By applying the same numerical techniques, we plot the conductivity in Figure \ref{fig4}.
It should be emphasized that due to the regularity condition at the horizon,
the resulting retarded Green's function is real, which implies that the
conductivity is a pure imaginary number.

\begin{figure}
\begin{center}
\vspace{1cm}
\includegraphics[angle=0,width=0.5\textwidth]{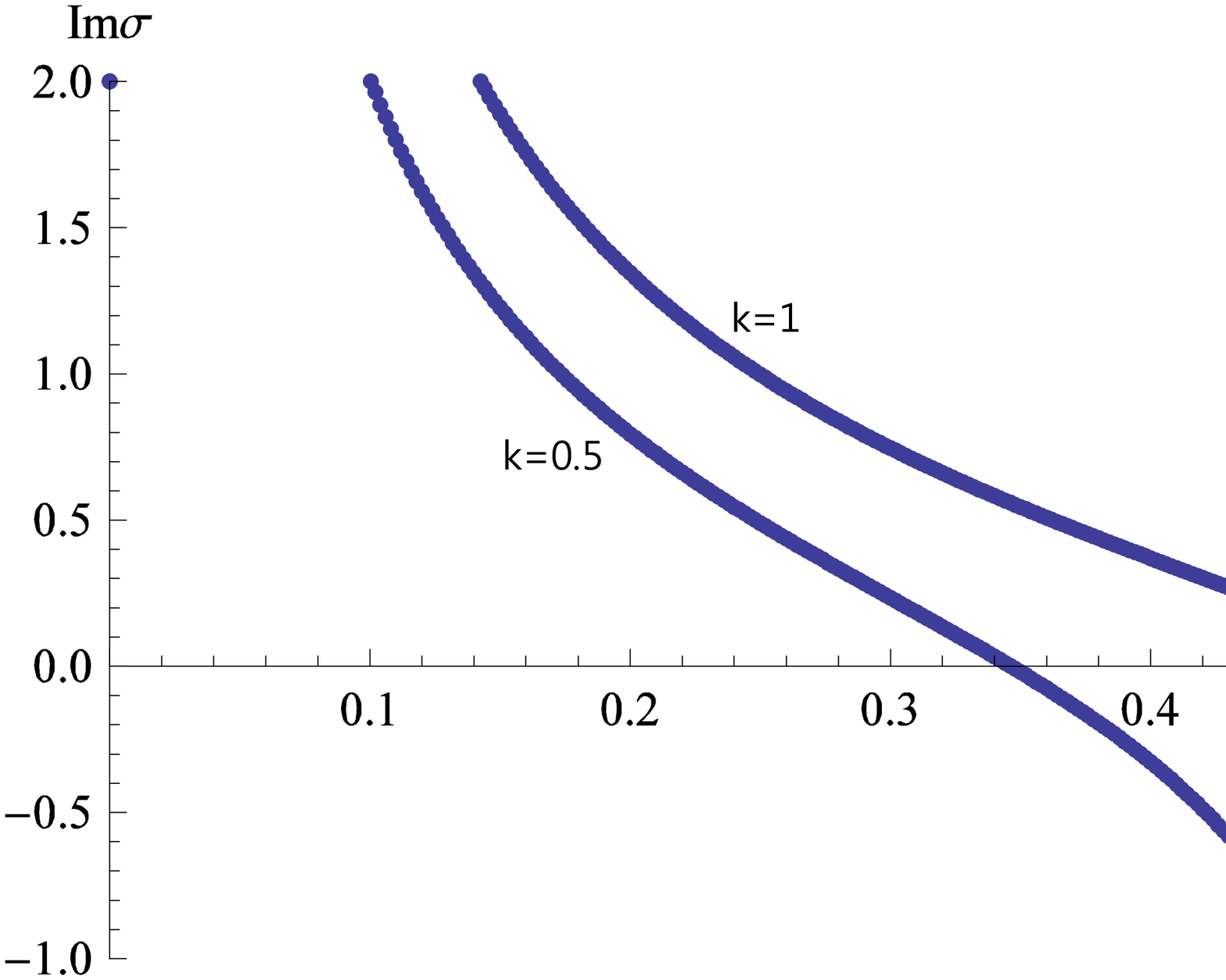}
\vspace{-2.5cm} \\
\caption{\small The imaginary conductivity at $a_0 = 1$ with $1/2 < a_1  < b_1 \leq 1$.}
\label{fig4}
\end{center}
\end{figure}
\item $a_1>1$\\
In this case, (\ref{eq55}) becomes
\begin{equation}
0 = a_x'' - \frac{2 (a_1 - 1)}{z}  a_x' + ( \frac{\omega^2}{a_0^4} z^{4(a_1-1)}
-\frac{k^2}{a_0^2 } z^{2 (a_1 + b_1 -2)} ) a_x .
\end{equation}
At the horizon($z \rightarrow\infty$), the $\omega^2$ term dominates, so the
approximate solution satisfying the incoming boundary condition is given by
\begin{equation}
a_x \approx  \ \exp (\frac{i \omega z^{2 a_1 -1}  }{a_0^2 (2 a_1 - 1)}) .
\end{equation}
Near the boundary ($z \rightarrow 0$), the $k^2$ term dominates,
so the approximate solution becomes
\begin{equation}
a_x \approx a_0 ( 1 + c \ z^{2 a_1 -1}),
\end{equation}
where $a_0$ is the boundary value of $a_x$. Then we have to perform similar numerical evaluations
to obtain the conductivity.

In Figure \ref{zeroa1beb1}, we plot several specific examples of conductivity and exhibit their
dependence on the momentum, which are very similar to those obtained in the $1/2 < b_1 < a_1 \leq 1 $
case. Notice that for $1/2< b_1 < a_1 \leq 1 $ and $a_1 >1$ the DC conductivity
is zero at $k = 1$ or $k=2$.
\begin{figure}
\begin{center}
\vspace{1cm}
\hspace{-0.5cm}
\includegraphics[angle=0,width=0.5\textwidth]{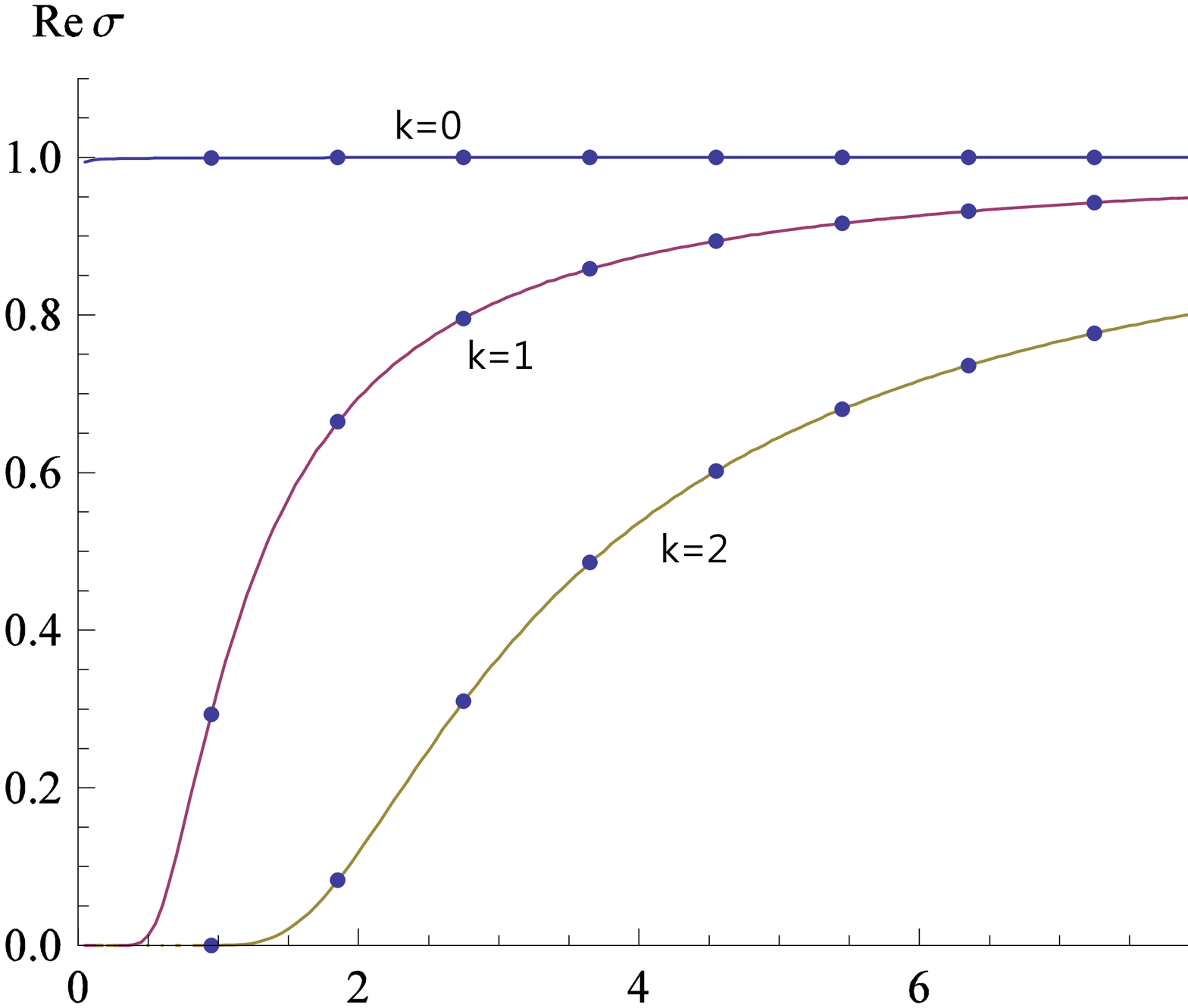}
\hspace{-0.5cm}
\includegraphics[angle=0,width=0.5\textwidth]{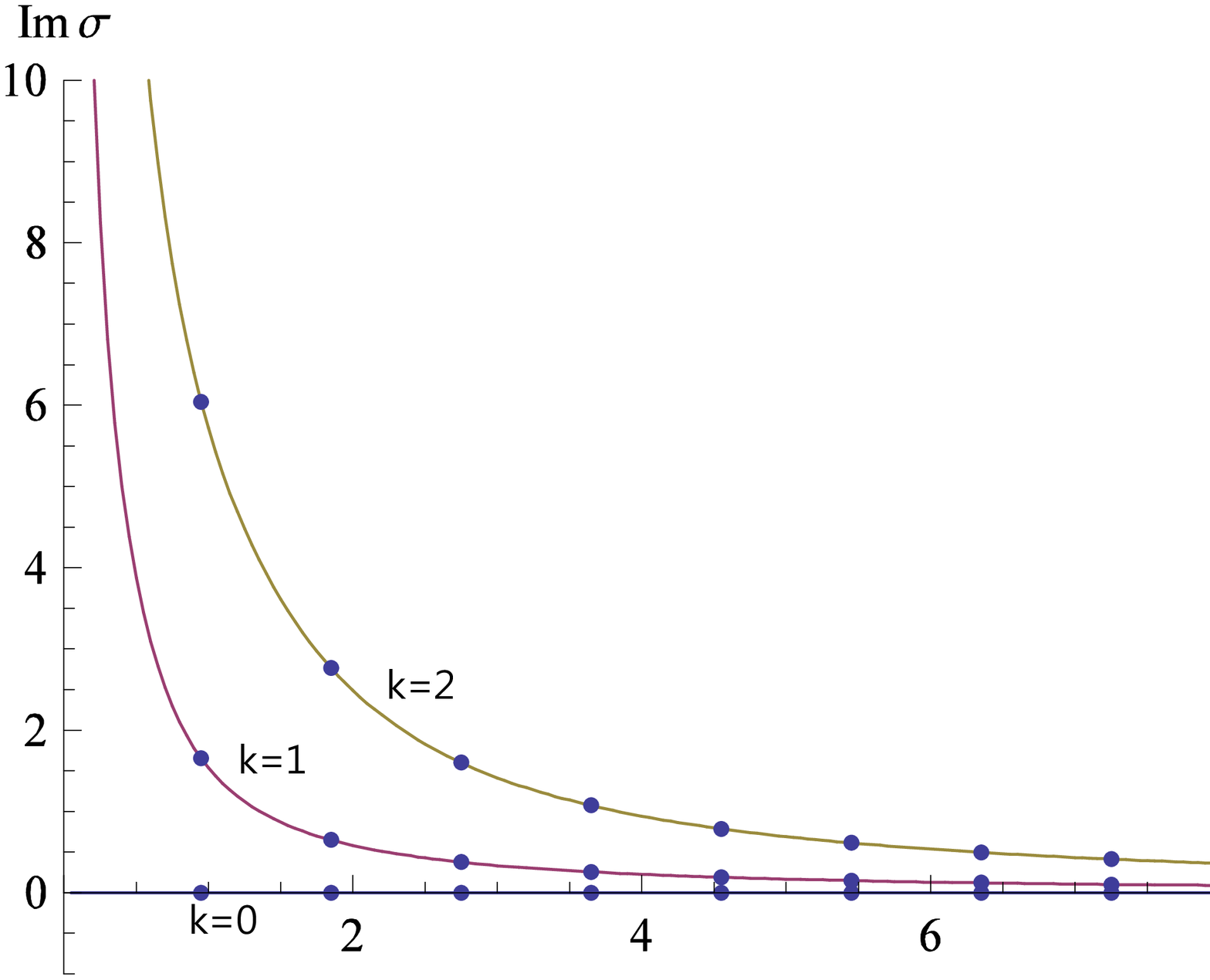}
\vspace{-2.5cm} \\
\caption{\small The real and imaginary conductivity when $a_1 = 5/4$, $b_1 = 3/4$ and $a_0 = 1$ .}
\label{zeroa1beb1}
\end{center}
\end{figure}
\end{itemize}
\subsubsection{At finite temperature}
First, we consider the simplest case $k=0$ in which we can obtain an exact solution.
After introducing a new coordinate
\begin{equation}
d u = \frac{dz}{z^{2 - 2 a_1} f},
\end{equation}
the Maxwell equation~(\ref{eq55}) for $k=0$ reduces to
\begin{equation}
\label{eq77}
0 = \partial_u^2 a_x + \frac{\omega^2}{a_0^4} a_x ,
\end{equation}
At the horizon, the solution to~(\ref{eq77}) satisfying the incoming boundary condition
is given by
\begin{equation}
a_x = a_0 \exp ( i \frac{\omega u}{a_0^2}) ,
\end{equation}
where $a_0$ is the boundary value of $a_x$. Near the boundary, the expansion of the
above solution becomes
\begin{equation}
a_x =a_0 (1 + i \frac{\omega}{(2 a_1 -1) a_0^2} z^{2 a_1 -1}) ,
\end{equation}
which is the same as one in the zero temperature case. Therefore, the retarded Green's function
and the conductivity at $k=0$ is the same as those in the zero temperature case.

\begin{figure}
\begin{center}
\vspace{1cm}
\hspace{-0.5cm}
\includegraphics[angle=0,width=0.5\textwidth]{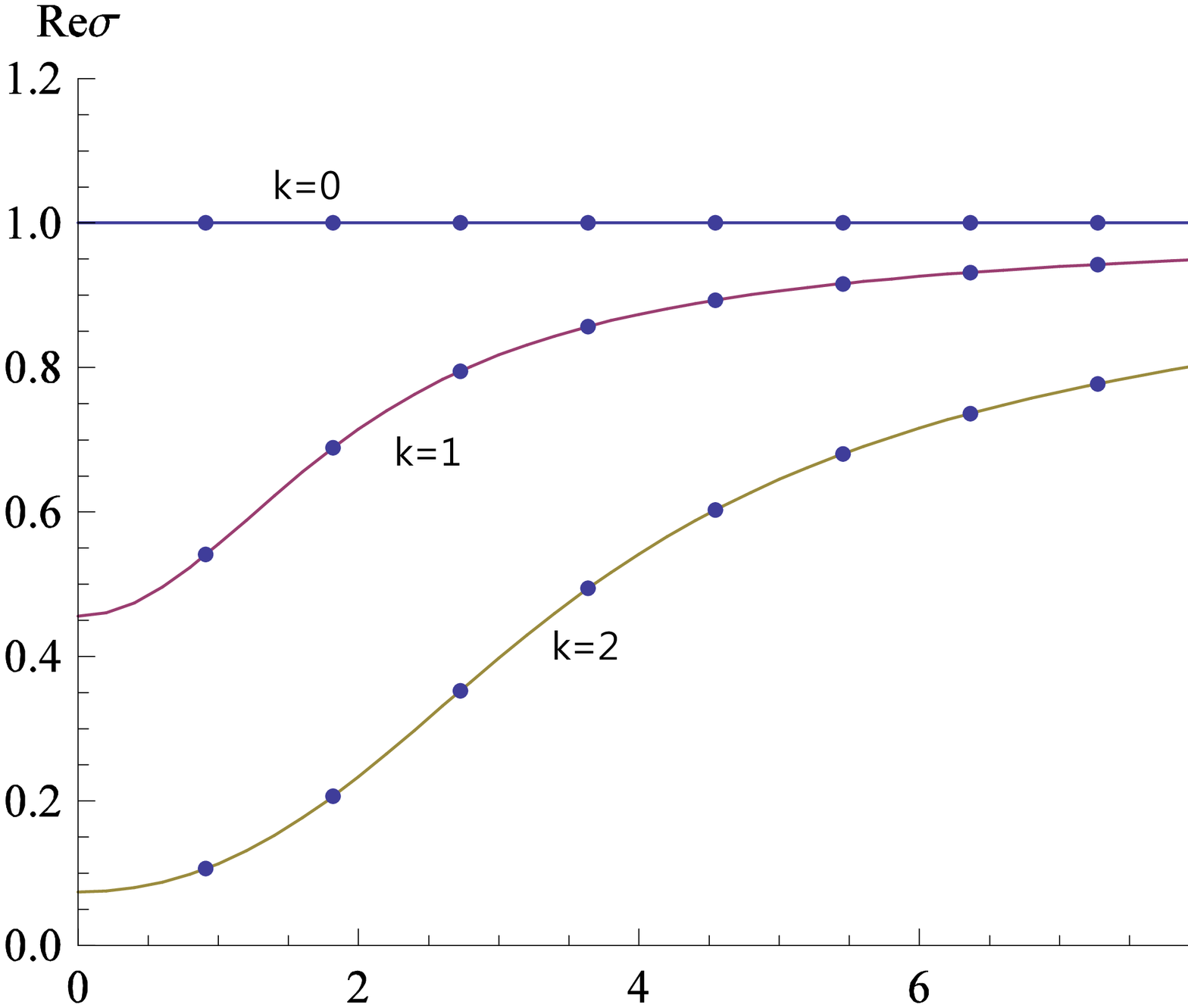}
\hspace{-0.5cm}
\includegraphics[angle=0,width=0.5\textwidth]{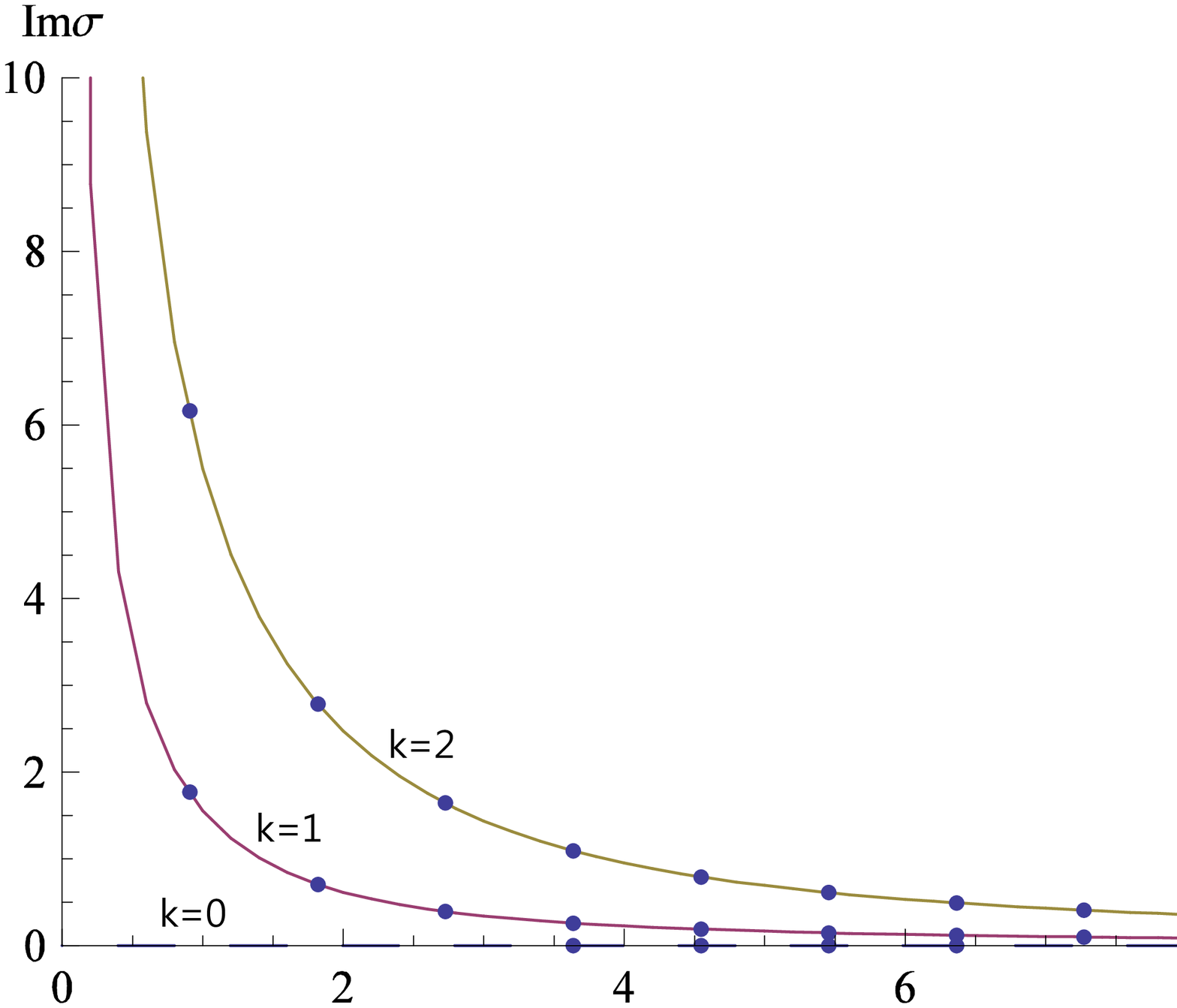}
\vspace{-2.5cm} \\
\caption{\small The real and imaginary conductivity at finite temperature $T = 0.239$ ($z_+=1$)
when $a_1 = 5/4$, $b_1 = 3/4$ and $a_0 = 1$ .}
\label{fig6}
\end{center}
\end{figure}

\begin{figure}
\begin{center}
\vspace{1cm}
\hspace{-0.5cm}
\includegraphics[angle=0,width=0.5\textwidth]{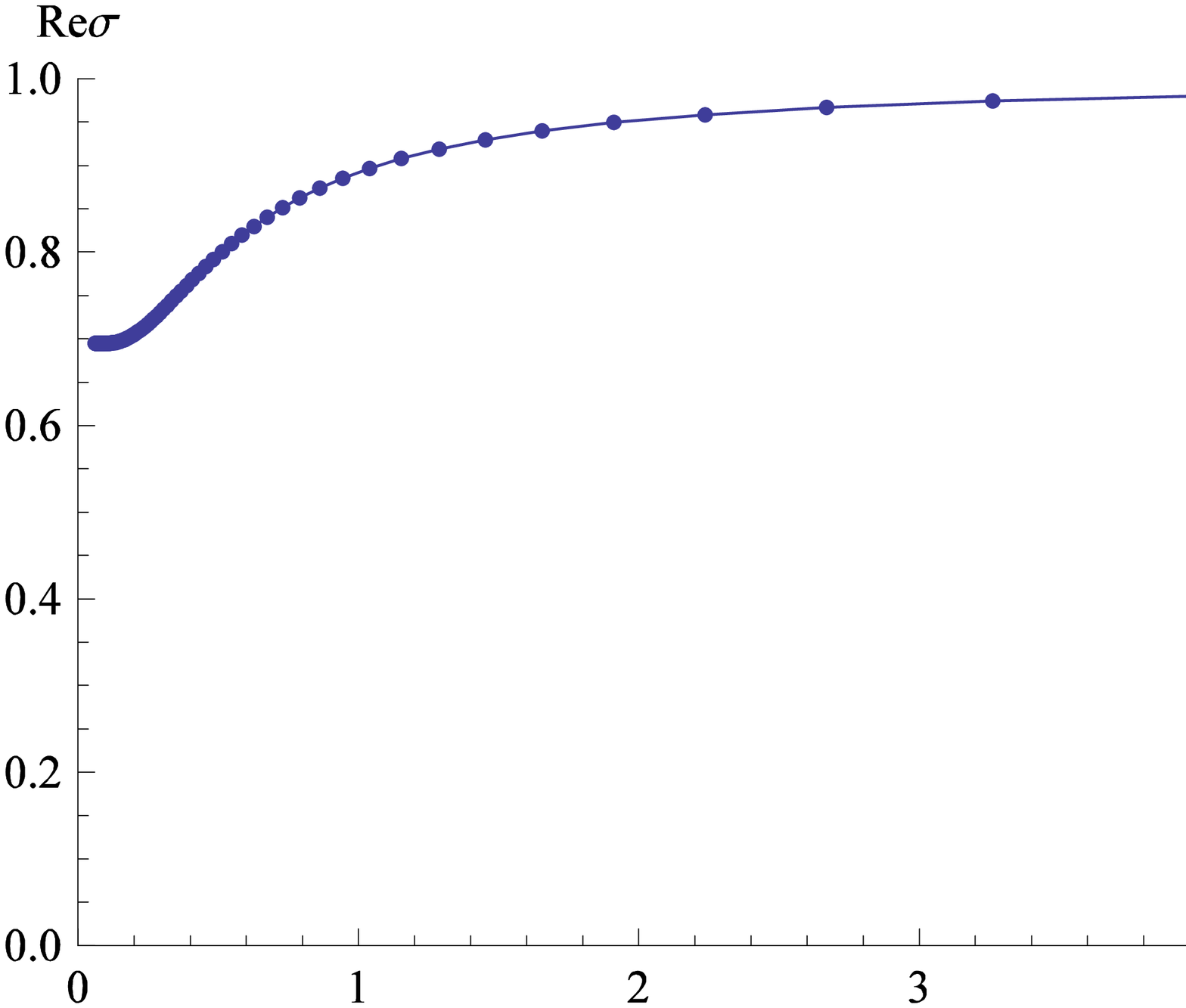}
\hspace{-0.5cm}
\includegraphics[angle=0,width=0.5\textwidth]{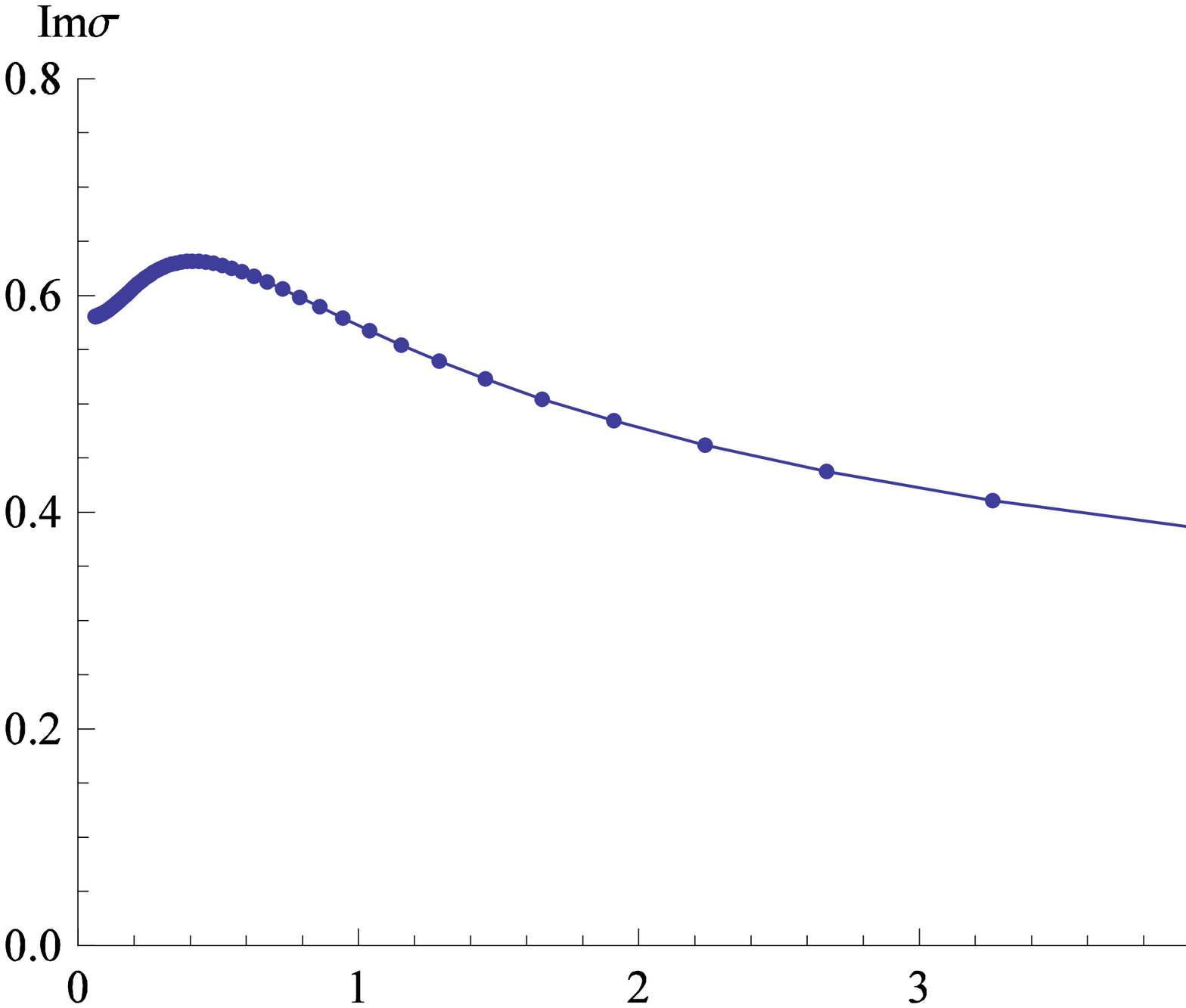}
\vspace{-2.5cm} \\
\caption{\small The dependence on temperature of the real and imaginary conductivity at
$\omega=2$, $k=1$, $a_1 = 5/4$, $b_1 = 3/4$ and $a_0 = 1$ .}
\label{fig7}
\end{center}
\end{figure}

Next we consider the general case with non-zero $k$. In this case, it is very difficult
to find an analytic solution, so we have to resort to numerical techniques which have been
frequently adopted in previous calculations. Figure \ref{fig6} shows the conductivity at finite temperature.
It can be seen that similar to the zero temperature cases, at $k=0$ the real part of the conductivity is still
a constant. But at $k=1$ or $k=2$, the conductivity at finite temperature goes to a constant
as the frequency goes to zero, while the conductivity at zero temperature approaches to zero.
This implies that the finite temperature DC conductivity is a constant while
the zero temperature DC conductivity is zero. We can also see that like the zero
temperature case, the conductivity grows as the frequency increases,
which is different from the strange metallic behavior.
Moreover, since our background geometry
is not a maximally symmetric space, the dual boundary theory is not conformal. So we can
expect that there exists non-trivial temperature dependence of the conductivity.
In Figure \ref{fig7} we plot the dependence on temperature of the electric conductivity.

\section{Strange metallic behavior from D-branes}
In the previous section we discussed how to obtain the conductivity via the conventional approach.
However, a complementary approach, which involves dynamics of probe D-branes, was proposed
in~\cite{Karch:2007pd}. Moreover, the authors of~\cite{Hartnoll:2009ns} treated probe D-branes
in four-dimensional Lifshitz black hole background as charge carriers and calculated the conductivities
by employing this complementary approach. In this section we will review the main contents of
our work~\cite{Lee:2010ii}.
\subsection{The main techniques}
In the beginning we give a brief review of the main techniques which were employed in~\cite{Karch:2007pd}
and~\cite{O'Bannon:2007in}. Without loss of generality, we take Dp-Dq brane system, namely
$N_f$ probe Dq-branes in the background of $N_c$ Dp-branes with $p<7$. Generally, the induced
Dq-brane metric can be expressed as follows
\begin{equation}
ds^{2}_{D_{q}}=g_{zz}dz^{2}+g_{tt}dt^{2}+g_{xx}d\vec{x}^{2}+g_{SS}d\Omega^{2}_{n}.
\end{equation}
In the above configuration we have assumed that the metric depends on the radial coordinate $z$
only. The probe Dq-branes wrap an $n$-dimensional sphere $S^{n}$ transverse to the Dp-brane
worldvolumes, where $g_{SS}$ denotes the metric component on the sphere. The Dq-branes also
possess $d=q-n-1$ directions parallel to the Dp-branes. Such a system may describe a holographic
defect, depending on the explicit values of $p$ and $q$. We assume that there exists a horizon
$z_{H}$ on the Dq-brane worldvolumes which locates at $g_{tt}(z_{H})=0$. Moreover, the Dp-brane
background also involves a nontrivial dilaton field $\phi(z)$.

The dynamics of the probe Dq-branes is described by the Dirac-Born-Infeld (DBI) action
\begin{equation}
S_{Dq}=-N_{f}T_{Dq}\int d^{8}\xi\sqrt{-{\rm det}(g_{ab}+2\pi\alpha^{\prime}F_{ab})}.
\end{equation}
Here $\xi$ denotes the coordinates on the Dq-brane worldvolume and $T_{Dq}$ is the tension
of the probe Dq-branes, while $g_{ab}$ and $F_{ab}$ are the induced metric and worldvolume
$U(1)$ field strength. The nontrivial components of the worldvolume $U(1)$ gauge field are
given by
\begin{equation}
A_{t}=A_{t}(z),~~~A_{x}(z,t)=-Et+h(z).
\end{equation}
Then the DBI action turns out to be
\begin{equation}
S_{Dq}=-N_{f}T_{Dq}V_{n}\int dzdte^{-\phi}g^{(d-1)/2}_{xx}g^{n/2}_{SS}
\sqrt{g_{zz}|g_{tt}|g_{xx}-(2\pi\alpha^{\prime})^{2}\left(g_{xx}A^{\prime2}_{t}+g_{zz}\dot{A}^{2}_{x}-
|g_{tt}|A^{\prime2}_{x}\right)},
\end{equation}
where $V_{n}$ denotes the volume of a unit $S^{n}$ and dot and prime denote partial derivatives
with respect to $t$ and $z$. Note that we have divided the volume of $R^{d}$ on both sides.

Since the action only involves z-derivatives of $A_{t}$ and $A_{x}$, there exist two conserved
quantities associated with them respectively. The conserved quantity associated with $A_{t}$
is given by
\begin{equation}
\frac{-\mathcal{N}_{q}(2\pi\alpha^{\prime})^{2}g_{xx}e^{-\phi}g^{(d-1)/2}_{xx}g^{n/2}_{SS}A^{\prime}_{t}}
{\sqrt{g_{zz}|g_{tt}|g_{xx}-(2\pi\alpha^{\prime})^{2}\left(g_{xx}A^{\prime2}_{t}+g_{zz}\dot{A}^{2}_{x}-
|g_{tt}|A^{\prime2}_{x}\right)}}=D,
\end{equation}
while the one associated with $A_{x}$ is
\begin{equation}
\frac{\mathcal{N}_{q}(2\pi\alpha^{\prime})^{2}|g_{tt}|e^{-\phi}g^{(d-1)/2}_{xx}g^{n/2}_{SS}h^{\prime}(z)}
{\sqrt{g_{zz}|g_{tt}|g_{xx}-(2\pi\alpha^{\prime})^{2}\left(g_{xx}A^{\prime2}_{t}+g_{zz}\dot{A}^{2}_{x}-
|g_{tt}|A^{\prime2}_{x}\right)}}=B.
\end{equation}
Then one can express $A^{\prime}_{t}(z)$ and $h^{\prime}(z)$ in terms of $E, D$ and $B$ after some algebra.
Furthermore, the on-shell DBI action turns out to be
\begin{equation}
\label{4eq7}
S_{Dq}\propto-\mathcal{N}_{q}\int dzdt\sqrt{\frac{|g_{tt}|g_{xx}-(2\pi\alpha^{\prime})^{2}E^{2}}
{e^{-2\phi}g_{xx}^{d}g^{n}_{SS}|g_{tt}|+\frac{|g_{tt}|D^{2}-g_{xx}B^{2}}
{\mathcal{N}_{q}^{2}(2\pi\alpha^{\prime})^{2}}}},
\end{equation}
where certain positive factors in the on-shell action have been omitted.

It can be seen that both the numerator and denominator under the square root
of~(\ref{4eq7}) are positive at the boundary $z=0$, while both of them are negative
at the horizon $z=z_{H}$. Therefore if we require that $S_{Dq}$ always remains
real from the horizon to the boundary, both the numerator and the denominator
should change sign at the same point $z_{\ast}$ with $0\leqslant z_{\ast}\leqslant z_{H}$.
Such a requirement imposes the following constraints
\begin{equation}
 |g_{tt}|g_{xx}-(2\pi\alpha^{\prime})^{2}E^{2}=0,
\end{equation}
\begin{equation}
e^{-2\phi}g_{xx}^{d}g^{n}_{SS}|g_{tt}|+\frac{|g_{tt}|D^{2}-g_{xx}B^{2}}
{\mathcal{N}_{q}^{2}(2\pi\alpha^{\prime})^{2}}=0,
\end{equation}
where all the metric quantities are evaluated at $z_{\ast}$.

After imposing appropriate boundary conditions for $A_{t}$ and $A_{x}$, one can
arrive at the following identifications\footnote{For details see~\cite{Karch:2007pd}.}
\begin{equation}
\langle J^{t}\rangle=D,~~~\langle J^{x}\rangle=B.
\end{equation}
Finally the conductivity is given by Ohm's law $\sigma=\langle J^{x}\rangle/E$
\begin{equation}
\sigma=\sqrt{\mathcal{N}^{2}_{q}(2\pi\alpha^{\prime})^{4}e^{-2\phi}g^{d-2}_{xx}g^{n}_{SS}
+(2\pi\alpha^{\prime})^{2}g^{-2}_{xx}\langle J^{t}\rangle^{2}}.
\end{equation}

The Hall conductivity can be evaluated in a similar way, which was illustrated in~\cite{O'Bannon:2007in}.
In this case the components of the worldvolume $U(1)$ gauge field are
\begin{equation}
A_{t}=A_{t}(z),~~~A_{x}(z,t)=-Et+f_{x}(z),~~~A_{y}(z,x)=Bx+f_{y}(z),
\end{equation}
Hence the on-shell DBI action can be expressed as
\begin{equation}
S_{Dq}\propto\int dzdt\frac{\xi}{\sqrt{\xi\chi-a^{2}}},
\end{equation}
where
\begin{equation}
\xi=|g_{tt}|g^{2}_{xx}+(2\pi\alpha^{\prime})^{2}(|g_{tt}|B^{2}-g_{xx}E^{2}),
\end{equation}
\begin{equation}
a=(2\pi\alpha^{\prime})^{2}(|g_{tt}|\langle J^{t}\rangle B+g_{xx}\langle J^{y}\rangle E),
\end{equation}
and
\begin{equation}
\chi=|g_{tt}|g_{xx}^{2}c(z)^{2}+(2\pi\alpha^{\prime})^{2}(|g_{tt}|\langle J^{t}\rangle^{2}
-g_{xx}(\langle J^{x}\rangle^{2}+\langle J^{y}\rangle^{2})),
\end{equation}
\begin{equation}
c(z)=\mathcal{N}_{q}(2\pi\alpha^{\prime})^{2}e^{-\phi(z)}g^{d/2-1}_{xx}g^{n/2}_{SS},~~~
\mathcal{N}_{q}=N_{f}T_{Dq}V_{n},
\end{equation}

In this case, when we require that the on-shell action is always real between the horizon
$z=z_{H}$ and the boundary $z=0$, the only way is to impose $\xi=\chi=a=0$ simultaneously
at $z=z_{\ast}$. Finally by making use of the formula
$\langle J_{i}\rangle=\sigma_{ij}E_{j}$, we obtain
\begin{equation}
 \sigma_{xx}=\frac{g_{xx}}{g^{2}_{xx}+(2\pi\alpha^{\prime})^{2}B^{2}}
\sqrt{(g^{2}_{xx}+(2\pi\alpha^{\prime})^{2}B^{2})c(z_{\ast})^{2}+(2\pi\alpha^{\prime})^{2}\langle J^{t}\rangle^{2}},
\end{equation}
\begin{equation}
\sigma_{xy}=\frac{(2\pi\alpha^{\prime})^{2}\langle J^{t}\rangle B}{g^{2}_{xx}+(2\pi\alpha^{\prime})^{2}B^{2}}.
\end{equation}

The above mentioned approach is simple and straightforward, and it would be easy to
perform generalizations in other backgrounds. In the next subsection we will evaluate
the conductivity in our anisotropic background via this approach.

\subsection{Massless charge carriers}
As advocated in~\cite{Hartnoll:2009ns}, once we started to
investigate mechanisms for strange metal behaviors by holographic
technology, we should involve a sector of (generally massive) charge
carriers carrying nonzero charge density $J^{t}$, interacting with
themselves and with a larger set of neutral quantum critical degrees
of freedom. The role of charge carriers was played by probe D-branes,
either massless or massive. Here the word ``massive'' means that the probe
D-branes possess nontrivial embedding profiles when wrapping on the internal manifold,
while the embedding profiles for massless charge carriers are trivial.
In this subsection we focus on the case of massless charge
carriers and the massive case will be considered subsequently.

\subsubsection{DC conductivity}
After taking the following coordinate transformation
\begin{equation}
v=\frac{1}{r},~~~v_{+}=\frac{1}{r_{+}},
\end{equation}
the metric~(\ref{2eq10}) becomes
\begin{equation}
ds^{2}=-\frac{a_{0}^{2}}{v^{2a_{1}}}f(v)dt^{2}+\frac{v^{2a_{1}-4}}{a_{0}^{2}f(v)}dv^{2}
+\frac{1}{v^{2b_{1}}}(dx^{2}+dy^{2}),
\end{equation}
where
\begin{equation}
f(v)=1-\frac{v^{\delta}}{v^{\delta}_{+}},~~~\delta=2a_{1}+2b_{1}-1.
\end{equation}
The temperature of the black hole can be rephrased as
\begin{equation}
\label{3eq4} T=\frac{1}{4\pi}\delta a_{0}^{2}v_{+}^{1-2a_{1}}.
\end{equation}

Next we consider the probe Dq-brane whose dynamics is described by
the following DBI action
\begin{equation}
S_{q}=-T_{q}\int d\tau d^{q}\sigma e^{-\phi}\sqrt{-{\rm
det}(g_{ab}+2\pi\alpha^{\prime}F_{ab})},
\end{equation}
where $T_{q}=(g_{s}(2\pi)^{q}l_{s}^{q+1})^{-1}$ is the Dq-brane
tension. Notice that the Wess-Zumino terms are neglected here. Considering
the following embedding profile for the probe D-branes
\begin{equation}
\tau=t,~~~\sigma^{1}=x,~~~\sigma^{2}=y,~~~\sigma^{3}=v,~~~\{\sigma^{4},\cdots,\sigma^{q}\}=\Sigma,
\end{equation}
the DBI action can be rewritten as
\begin{equation}
\label{3eq7} S_{q}=-\tau_{\rm eff}\int
dtdvd^{2}xv^{-k_{0}}\sqrt{-{\rm
det}(g_{ab}+2\pi\alpha^{\prime}F_{ab})},
\end{equation}
where $\tau_{\rm eff}=T_{q}{\rm Vol}(\Sigma)$, ${\rm Vol}(\Sigma)$
denoting the volume of the compact manifold. Note that here the
dilaton has a non-trivial dependence on the radial coordinate $v$,
$e^{-\phi}=v^{-k_{0}}$. As emphasized in~\cite{Hartnoll:2009ns},
incorporating a non-trivial dilaton might lead to a more realistic
holographic model of strange metals and we will see that this is
indeed the case.

We take the following ansatz for the worldvolume $U(1)$ gauge field
\begin{equation}
A=\Phi(v)dt+(-Et+h(v))dx.
\end{equation}
Following~\cite{Karch:2007pd}, we can finally arrive at the DC conductivity
\begin{equation}
\label{3eq18} \sigma=\sqrt{(2\pi\alpha^{\prime})^{4}\tau_{\rm
eff}^{2}v^{-2k_{0}}_{\ast}+(\frac{2\pi\alpha^{\prime}}{L^{2}})^{2}(J^{t})^{2}v^{4b_{1}}_{\ast}}.
\end{equation}
There exist two terms in the square root. One may
interpret the first term as arising from thermally produced pairs of
charge carriers, though here it has some non-trivial dependence on
$v_{\ast}$. It is expected that such a term should be suppressed
when the charge carriers have large mass. Then the surviving term
gives
\begin{equation}
\sigma=\frac{2\pi\alpha^{\prime}}{L^{2}}J^{t}v^{2b_{1}}_{\ast}.
\end{equation}
By combining~(\ref{3eq4}), one can obtain the power-law for the DC
resistivity,
\begin{equation}
\rho\sim\frac{T^{\lambda}}{J^{t}},~~~\lambda=\frac{2b_{1}}{2a_{1}-1},
\end{equation}
where we take the limit $E\ll1$ so that $v_{\ast}\approx v_{+}$.
Note that when the parameter $\eta$ in the Liouville
potential is zero, the background reduces to a Lifshitz-like solution
at finite temperature with $a_{1}=1$ and $b_{1}=z^{-1}$.
Thus we have $\rho\sim T^{2/z}/J^{t}$, which agrees
with the result obtained in~\cite{Hartnoll:2009ns}.
\subsubsection{DC Hall conductivity}
As reviewed in previous subsection, similarly here we
can calculate the Hall conductivity in our anisotropic background.
Here we take the following ansatz for the worldvolume $U(1)$ gauge
fields
\begin{equation}
A_{t}=\Phi(v),~~~A_{x}(v,t)=-Et+f_{x}(v),~~~A_{y}(v,x)=Bx+f_{y}(v).
\end{equation}
Following~\cite{O'Bannon:2007in} we can finally obtain
\begin{equation}
\sigma^{xx}=\frac{v^{-k_{0}}_{\ast}}{1+(\frac{2\pi\alpha^{\prime}}{L^{2}})^{2}B^{2}v^{4b_{1}}_{\ast}}
\sqrt{(2\pi\alpha^{\prime})^{4}\tau_{\rm
eff}^{2}(1+(\frac{2\pi\alpha^{\prime}}{L^{2}})^{2}B^{2}v^{4b_{1}}_{\ast})+(
\frac{2\pi\alpha^{\prime}}{L^{2}})^{2}(J^{t})^{2}
v^{4b_{1}+2k_{0}}_{\ast}},
\end{equation}
and
\begin{equation}
\sigma^{xy}=\frac{(2\pi\alpha^{\prime})^{2}BJ^{t}v^{4b_{1}}_{\ast}}
{L^{4}+(2\pi\alpha^{\prime})^{2}B^{2}v^{4b_{1}}_{\ast}}.
\end{equation}
Here are some remarks on the results for the conductivity tensor.
\begin{itemize}
\item When both $B$ and $E$ are small, the Hall conductivity
becomes $\sigma^{xy}\sim T^{4b_{1}/(1-2a_{1})}$. Once we take
$\eta=0$ in the Liouville potential, we have $a_{1}=1$ and therefore
$\sigma^{xy}\sim T^{-4b_{1}}$. Note that $b_{1}=1/z$, so we recover
the result obtained in~\cite{Hartnoll:2009ns} $\sigma^{xy}\sim
T^{-4/z}$.
\item The expression for $\sigma^{xx}$ reduces to the one obtained
in previous subsection when $B=0$. Furthermore, when the second term
in the square root dominates, and $B$ is small, we reproduce the
result $\sigma^{xx}\sim T^{-\lambda}$ where
$\lambda=\frac{2b_{1}}{2a_{1}-1}$.
\item One interesting quantity for studying the strange metals is
the ratio $\sigma^{xx}/\sigma^{xy}$. When the first term in the
square root of $\sigma^{xx}$ is subdominant, one can easily obtain
the following result
\begin{equation}
\frac{\sigma^{xx}}{\sigma^{xy}}\sim
v^{-2b_{1}}_{\ast}=T^{\frac{-2b_{1}}{1-2a_{1}}}.
\end{equation}
In the $\eta=0$ limit, $a_{1}=1, b_{1}=1/z$, we have
$$\frac{\sigma^{xx}}{\sigma^{xy}}\sim
v^{-2/z}_{\ast}=T^{2/z}.$$
\item The strange metals exhibit the following anomalous behaviors:
$\sigma^{xx}\sim T^{-1}, \sigma^{xx}/\sigma^{xy}\sim T^{2}$. In
contrast, $\sigma^{xx}/\sigma^{xy}\sim(\sigma^{xx})^{-1}$ in Drude
theory. Since $\sigma^{xx}\sim T^{-2/z}$ in the limit of $\eta=0$,
our result can mimic Drude theory in this limit, which agrees with
the result in~\cite{Hartnoll:2009ns}.
\end{itemize}
\subsubsection{AC conductivity}
The AC conductivity can be calculated by considering the fluctuations
of the probe gauge fields
\begin{equation}
\delta A=(A_{t}(v)dt+A_{x}(v)dx+A_{y}(v)dy)e^{-i(\omega t-kx)}.
\end{equation}
Then the DBI action can be expanded as
\begin{eqnarray}
S_{q}&=&-\tau_{\rm eff}\int
dtd^{2}xdvv^{-k_{0}}\gamma^{-1}\sqrt{-g_{tt}g_{vv}}g_{xx}[1+\frac{1}{2}\gamma
(2\pi\alpha^{\prime})^{2}\frac{F^{2}_{xy}}{g^{2}_{xx}}\nonumber\\& &
+\frac{1}{2}\gamma^{2}
(2\pi\alpha^{\prime})^{2}\frac{F^{2}_{iv}}{g_{vv}g_{xx}}-\frac{1}{2}\gamma^{2}
(2\pi\alpha^{\prime})^{2}\frac{F^{2}_{ti}}{g_{tt}g_{xx}}-\frac{1}{2}\gamma^{3}
(2\pi\alpha^{\prime})^{2}\frac{F^{2}_{tv}}{g_{tt}g_{vv}}],
\end{eqnarray}
Let us focus on the quadratic terms of $F_{\mu\nu}$
\begin{eqnarray}
S_{F}&=&-\frac{\tau_{\rm eff}}{2}(2\pi\alpha^{\prime})^{2}\int
dtd^{2}xdvv^{-k_{0}}\gamma[\frac{\sqrt{-g_{tt}g_{vv}}}{g_{xx}\gamma}
F^{2}_{xy}\nonumber\\& &+\sqrt{\frac{-g_{tt}}{g_{vv}}}F^{2}_{iv}
-\sqrt{\frac{-g_{vv}}{g_{tt}}}F^{2}_{ti}-\frac{\gamma
g_{xx}}{\sqrt{-g_{tt}g_{vv}}}F^{2}_{tv}],
\end{eqnarray}
from which we can derive the equation of motion for $A_{x}$
\begin{equation}
\partial_{v}(v^{-k_{0}}\sqrt{\frac{-g_{tt}}{g_{vv}}}\gamma A_{x}^{\prime})
=-v^{-k_{0}}\sqrt{\frac{g_{vv}}{-g_{tt}}}\gamma\omega^{2}A_{x},
\end{equation}
It was pointed out in~\cite{Gubser:2008wz, Horowitz:2009ij}
that the equation of motion for the fluctuation can be converted
into a Schr\"{o}dinger equation. For our case this can be realized
by defining
\begin{equation}
A_{x}=(v^{-k_{0}}\gamma)^{-1/2}\Psi,~~~\frac{d}{dv}=\frac{v^{2a_{1}-2}}{a^{2}_{0}f(v)}\frac{d}{ds}.
\end{equation}
Then the equation of motion for $A_x$ becomes
\begin{equation}
-\frac{d^{2}}{ds^{2}}\Psi+U\Psi=\omega^{2}\Psi,
\end{equation}
with the following effective potential
\begin{equation}
U=\frac{1}{2}\frac{1}{\sqrt{v^{-k_{0}}\gamma}}\frac{d}{ds}[\frac{1}{\sqrt{v^{-k_{0}}\gamma}}
\frac{d}{ds}(v^{-k_{0}}\gamma)].
\end{equation}

\begin{figure} \label{fig1}
\begin{center}
\vspace{3cm} \hspace{-0.5cm}
\includegraphics[angle=0,width=0.5\textwidth]{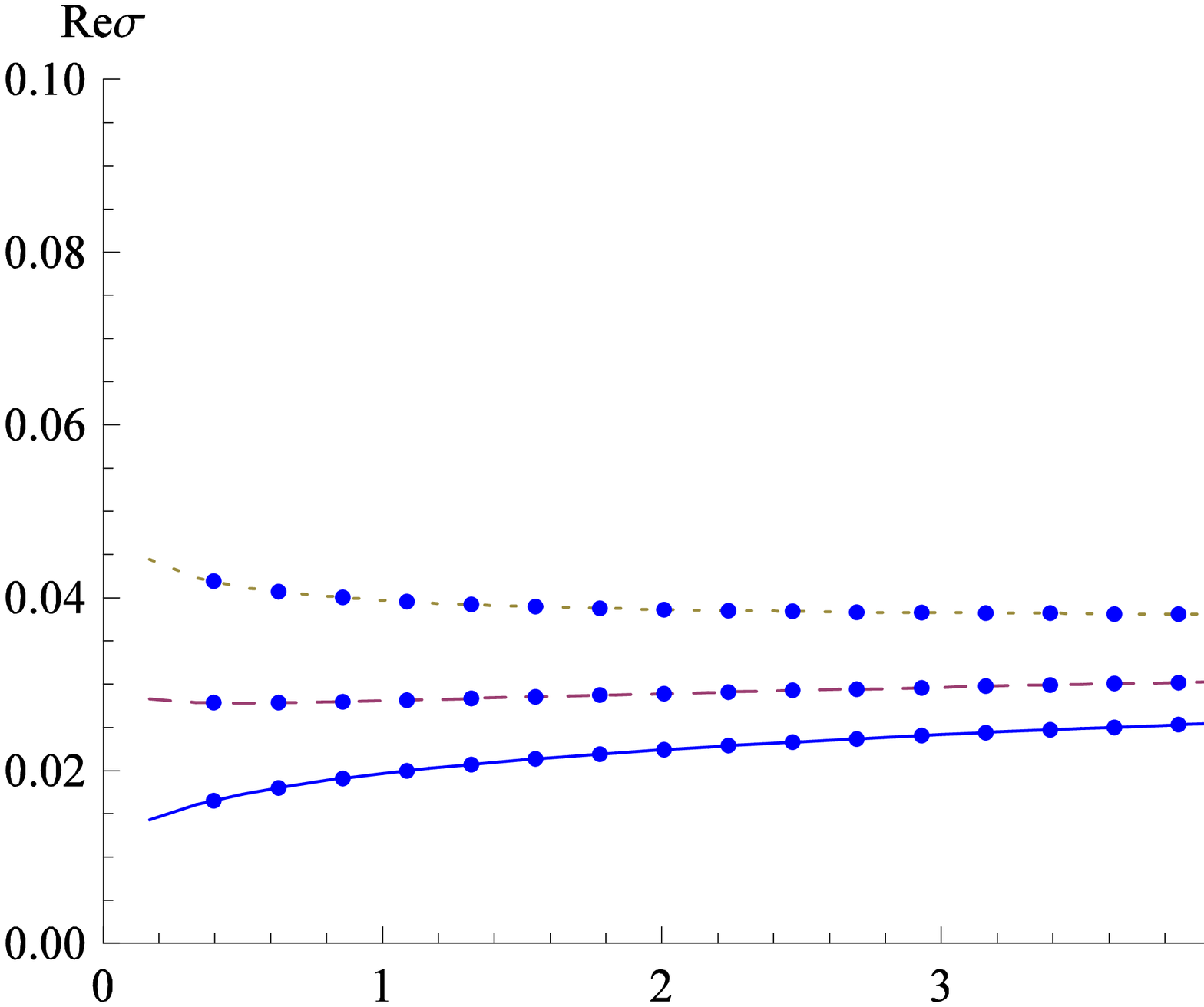}
\hspace{-0.5cm}
\includegraphics[angle=0,width=0.5\textwidth]{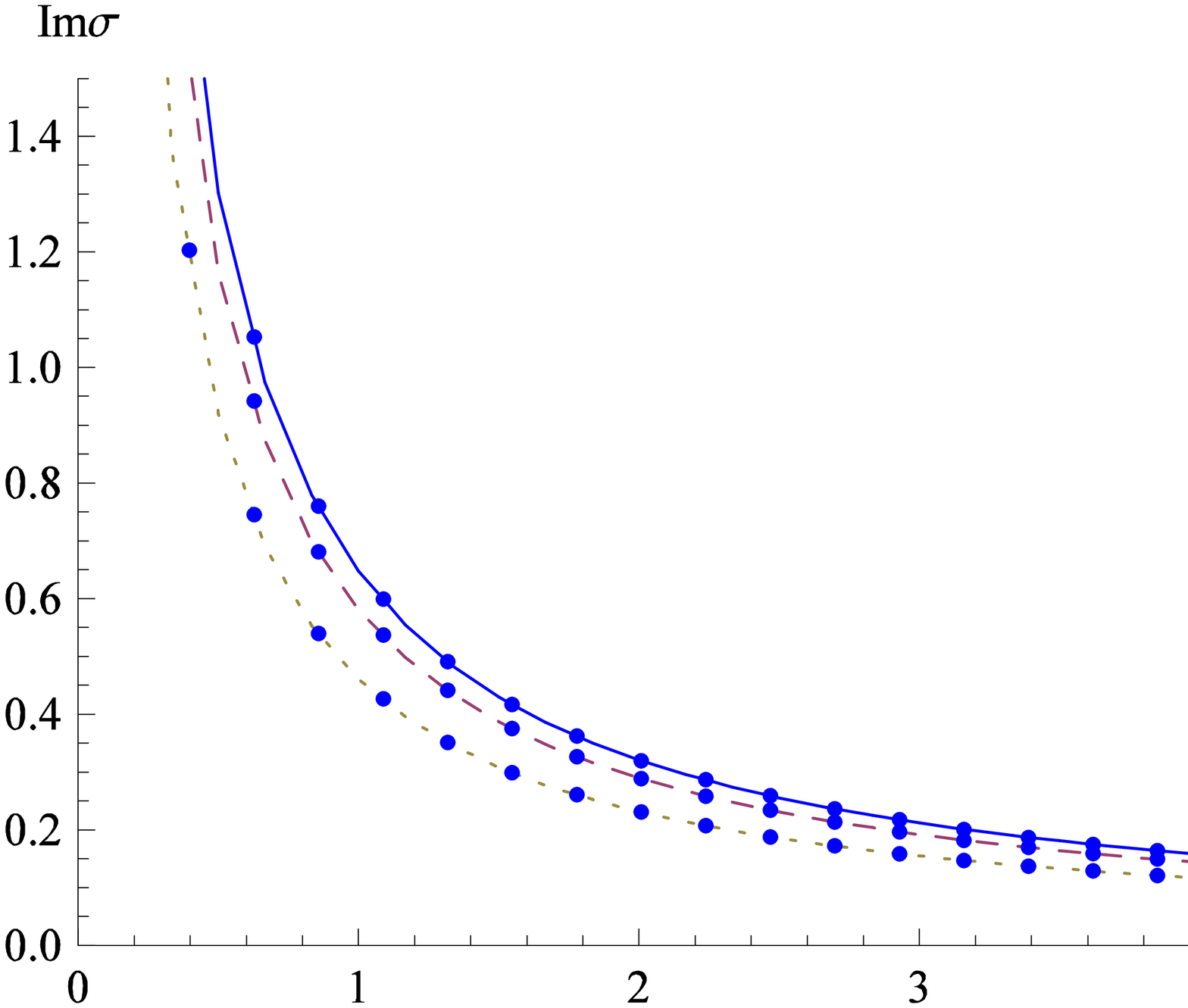}
\vspace{-2.5cm} \\
\caption{\small The real and imaginary conductivity for
$C=0$(solid), $2$(dashed) and $4$(dotted), where we set $2 \pi
\alpha^{\prime}=1$, $\alpha =-\eta =1$, $\Lambda = - 3$, $\tau_{\rm eff} = 1$ and $v_+
= 10$.}
\end{center}
\end{figure}

The AC conductivity can be evaluated by numerical methods, whose details will
be omitted here. The final results are plotted in Fig. \ref{fig1}, where we show the
real and imaginary conductivity depending on the charge density $C
\sim J^t$. It can be seen from the figure that at given $\omega$, the real and
imaginary conductivity increases and decreases respectively, as the
charge density increases.

Interestingly, even for zero density in Fig. \ref{fig1} there exists non-zero conductivity, which
may be related to the effect of the pair creation of charged particles. Usually,
as the energy goes up more charged particles can be created which can explain increasing of the
real conductivity at the high energy region. Another interesting point is that at the high density
and low frequency regime (see the case for $C=4$) the conductivity decreases as the frequency
increases. This aspect would be explained as the follow: in this regime the pair creation
of the charged particles generates induced electric field which diminishes
the background electric field. Since the charged carrier moves slowly due to the weakened
electric field, the amount of the charged carrier current also grows smaller. This
can explain the drop of the real conductivity in the high density and small energy regime.
In the large energy case above the some critical frequency,
the effect of the pair creation of charged particles would be more dominant, so
it makes the real conductivity increases as the energy increases.

\subsection{Massive charge carriers}
We will consider the effects of massive charge carriers in this
section, which are more closely related to model building for real
-world strange metals. In this case, the energy gap is large
compared to the temperature: $E_{\rm gap}\gg T$. When translated
into the language of probe D-branes, the massive charge carriers
correspond to flavor branes wrapping internal cycles, whose volumes
vary with radial direction~\cite{Karch:2002sh, Kobayashi:2006sb}.

As pointed out in~\cite{Karch:2002sh, Kobayashi:2006sb}, at
finite temperature the flavor brane shrinks to a point at $v=v_{0}$
for large enough mass. In this case the charge carriers correspond
to strings stretching from the flavor brane from $v=v_{0}$ to the
black hole horizon $v=v_{+}$. In a dilute limit one can consider a
small density of such strings and ignore the backreaction. For
larger densities the backreaction cannot be neglected and the
resulting configuration is that the brane forms a ``spike'' in place
of the strings. We will study the dilute limit first and then take
the backreaction into account.

\subsubsection{Conductivity in the dilute regime}
The massive charge carriers can be treated as strings stretching
from the probe brane to the black hole horizon in the dilute limit,
while there are no interactions between the strings. In this limit
the probe brane is described by the zero-density solution, which
has a cigar shape with $v=v_{0}$ at the tip.

Let us take the static gauge $t=\tau, v=\sigma$. When expanded to
quadratic order for the transverse fluctuations, the Nambu-Goto
action
\begin{equation}
S_{\rm NG}=-T\int d\tau d\sigma\sqrt{-{\rm det}h_{ab}}+\int A
\end{equation}
gives the following field equation
\begin{equation}
-\frac{1}{\alpha^{\prime}}\frac{a^{2}_{0}f(v)}{v^{2a_{1}+2b_{1}-2}}
\partial_{v}x^{i}+F_{i0}+F_{ij}\dot{x}^{j}=0.
\end{equation}
In the zero-frequency limit, the field equation can be easily
integrated out
\begin{equation}
x^{i}=V^{i}(t+ \frac{1}{v_+^{2 b_1}} \int^{v}\frac{u^{2a_{1}+2b_{1}-2}}{a_{0}^{2}f(u)}du),
\end{equation}
where $V^{i}$ is an integration constant and the relative
normalization of the two terms is fixed by imposing incoming boundary
conditions at the horizon. By assuming  $v_{0}\ll v_{+}$, at the
boundary we can obtain
\begin{equation}
\label{4eq4} \frac{1}{v_+^{2 b_1}}
\frac{1}{\alpha^{\prime}}V^{i}=F_{i0}+F_{ij}V^{j},
\end{equation}
According to Drude's law we have the following relation
\begin{equation}
\frac{m}{\tau}\propto\frac{1}{\alpha^{\prime}}T^{\frac{2 b_1}{2 a_1 -1}},
\end{equation}
Therefore the DC conductivity in the dilute limit of the massive
charge carriers is given by
\begin{equation}
\sigma=\frac{\tau}{m}J^{t}\propto\frac{J^{t}}{T^{\frac{2 b_1}{2 a_1 -1}}}.
\end{equation}

Next we consider the AC case. The bulk equation of motion for $x(v,t)={\rm
Re}(X_{\omega}(v)e^{-i\omega t})$ reads
\begin{equation}
\partial_{v}(\frac{a_{0}^{2}f(v)}{v^{2a_{1}+2b_{1}-2}}\partial_{v}X_{\omega})
=-\omega^{2}\frac{v^{2a_{1}-2b_{1}-2}}{a^{2}_{0}f(v)}X_{\omega}.
\end{equation}
Now consider the case of zero magnetic field $F_{10}=E, F_{ij}=0$,
at the boundary $v=v_{0}$ the surface term gives
\begin{equation}
\frac{1}{\alpha^{\prime}}\frac{a^{2}_{0}f(v_{0})}{v^{2a_{1}+2b_{1}-2}_{0}}
\partial_{v}X_{\omega}(v_{0})=E.
\end{equation}
Then the conductivity can be evaluated as
\begin{equation}
\sigma=\frac{J^{t}V_{\omega}(v_{0})}{E}=\frac{i\omega
J^{t}X_{\omega}(v_{0})}{E}=\frac{i\omega\alpha^{\prime}J^{t}X_{\omega}
(v_{0})v^{2a_{1}+2b_{1}-2}}{a^{2}_{0}f(v_{0})\partial_{\omega}X(v_{0})},
\end{equation}

In sum, we have the following scaling behaviors for the resistivity
and conductivity in the dilute regime,
\begin{equation}
\rho=\frac{1}{\sigma}\sim T^{\frac{2b_{1}}{2b_{1}-2a_{1}}},~~~
\sigma(\omega)\sim\omega^{1-2\xi},
\end{equation}
where $\xi=(2a_{1}+2b_{1}-1)/(2a_{1}-2b_{1})$. Recall that in real-world
strange metals,
\begin{equation*}
\rho\sim T^{\nu_{1}},~~~ \sigma(\omega)\sim\omega^{-\nu_{2}},
\end{equation*}
where $\nu_{1}\approx1,\nu_{2}\approx0.65$. Therefore if we require our
dual gravity background to exhibit the same scaling behavior, we have to set
\begin{equation}
\alpha=\pm0.293491,~~~\eta=\pm1.45201.
\end{equation}
The consistency of the above choices for the parameters has been
verified.
\subsubsection{Conductivity at finite densities}
When the backreaction of the massive charge carriers cannot be
neglected, we should introduce an additional scalar field which
corresponds to the ``mass'' operator in the boundary field theory.
The volume of the internal cycle wrapped by the probe brane is
determined by this scalar field. In this case the DBI action becomes
\begin{equation}
S_{q}=-\tau_{\rm eff}\int
dtd^{2}xdvv^{-k_{0}}V(\theta)^{n}\sqrt{g_{xx}}\sqrt{-g_{tt}g_{xx}g_{\sigma\sigma}-(2\pi
\alpha^{\prime})^{2}(g_{\sigma\sigma}E^{2}+g_{xx}\Phi^{\prime2}+g_{tt}h^{\prime2})},
\end{equation}
where $V(\theta)^{n}$ denotes the volume of the $n$-dimensional
submanifold wrapped by the probe brane. Notice that here we have
introduced the worldvolume gauge field as follows
\begin{equation}
A=\Phi(v)dt+(-Et+h(v))dx
\end{equation}
and the induced metric component is given by
$g_{\sigma\sigma}=g_{vv}+\theta^{\prime2}$.

The calculations of the DC conductivity are almost
the same as the massless case, except for that there exists an additional equation
determining the background profile of the probe brane $\theta(v)$
\begin{equation}
\frac{d}{dv}[v^{-2k_{0}}V(\theta)^{2n}g^{3/2}_{xx}\sqrt{\frac{-g_{tt}}{g_{\sigma\sigma}}}
\theta^{\prime}GF^{1/2}]-\frac{d}{d\theta}[v^{-2k_{0}}V(\theta)^{2n}g^{3/2}_{xx}
\sqrt{-g_{tt}g_{\sigma\sigma}}GF^{1/2}]=0.
\end{equation}
The DC conductivity is given by
\begin{equation}
\sigma=\sqrt{(2\pi\alpha^{\prime})^{4}\tau_{\rm
eff}^{2}v^{-2k_{0}}_{\ast}V(\theta_{\ast})^{2n}+(\frac{2\pi\alpha^{\prime}}{L^{2}})
^{2}(J^{t})^{2}v^{4b_{1}}_{\ast}}.
\end{equation}
Note that in the massless limit $V(\theta)~\rightarrow~1$, the above
result reduces to the one obtained in previous
section~(\ref{3eq18}).

The calculations of the AC conductivity are also similar to the massless
case.
 In particular, the equation for the fluctuation can still be transformed into a
Schr\"{o}dinger form. 
At first sight one may consider that we can perform similar
calculations as for the massless case. However, one key point in the
calculations of~\cite{Hartnoll:2009ns} was that in a wide
range $v_{0}<v<v_{+}$, the profile of the embedding was
approximately constant. This behavior was confirmed by numerical
calculations and played a crucial role in simplifying the corresponding
Schr\"{o}dinger equation. Unfortunately, here we cannot find such a
constant behavior for the profile of the embedding. The profile
turns out to be singular in the near horizon region. Thus we cannot
simplify the Schr\"{o}dinger equation considerably to find the
analytic solutions. It might be related to the non-trivial dilaton
field in the DBI action and we expect that such singular behavior
may be cured in realistic D-brane configurations, which might enable
us to calculate the transport coefficients.
\section{Summary}
Strange metals are fascinating subjects in condensed matter physics, though
comprehensive theoretical understanding on their properties is still under
investigation. Since AdS/CFT correspondence provides a powerful framework for
analyzing dynamics of strongly coupled field theories, it is desirable to see
if AdS/CFT can shed light on issues related to strange metals. In this paper
we review our recent work on holographic strange metals~\cite{Lee:2010xx}
and~\cite{Lee:2010ii}. We consider the effects induced by the bulk Maxwell fields, the
additional $U(1)$ gauge fields, as well as the probe D-branes.

First, we study fluctuations of the background gauge field, which is
coupled with the dilaton field. Due to this non-trivial dilaton coupling, the conductivity of this
system depends on the frequency non-trivially. After choosing appropriate parameters, at both
zero and finite temperature we obtain the strange metal-like AC conductivity proportional to
the frequency with a negative exponent.

Second, we also investigate the effects of an extra $U(1)$ gauge field fluctuations without
dilaton coupling. We classify all possible conductivities either analytically or numerically,
according to the ranges of the parameters in the bulk theory. We find that the conductivities
for $k=0$ at zero and finite temperature become constant, due to the trivial gauge coupling
in the action for the additional $U(1)$ gauge field. This implies that to describe
frequency-dependent conductivities, it would be important to consider nontrivial gauge coupling
at least in the current approach. We also investigate the dependence of the conductivity on the
spatial momentum and the temperature. We find that as the spatial momentum increased,
the real part of the conductivity went down. In addition, we also find that the DC conductivity
at finite temperature becomes a non-zero constant while the one at zero temperature is zero in this set-up.

Third, we discuss dynamics of both massless and massive charge carriers,
which are represented by probe D-branes. For massless charge carriers, we obtain the DC conductivity and DC Hall
conductivity by applying the approach proposed in~\cite{Karch:2007pd} and~\cite{O'Bannon:2007in}.
The results can reproduce those obtained in~\cite{Hartnoll:2009ns} in certain specific limits.
We also calculate the AC conductivity by transforming the corresponding equation of motion into
the Schr\"{o}dinger equation. For massive charge carriers, the DC and AC conductivities are also obtained in the dilute limit. When the parameters in the action take the following values, $\alpha =\pm 0.293491$ and
$\eta=\pm 1.45201$, we can realize the experimental values of the resistivity and the AC conductivity
for strange metals simultaneously in the dual gravity side. We also obtain the DC conductivity at finite density. One thing we are not able to analyze is the AC conductivity at finite density, which may be due to the
singular behavior of the D-brane embedding profiles. We expect that such a difficulty
may be cured in realistic D-brane configurations.  

\bigskip \goodbreak \centerline{\bf Acknowledgments}
\noindent This work was supported by the National Research
Foundation of Korea(NRF) grant funded by the Korea government(MEST)
through the Center for Quantum Spacetime(CQUeST) of Sogang
University with grant number 2005-0049409. DWP would like to thank Institute of Theoretical Physics, Chinese Academy of Sciences for hospitality, where most of the work was done. DWP
acknowledges an FCT (Portuguese Science Foundation) grant.
This work was also funded by FCT through projects
CERN/FP/109276/2009 and PTDC/FIS/098962/2008. C. Park
was also supported by Basic Science Research Program through the
National Research Foundation of Korea(NRF) funded by the Ministry of
Education, Science and Technology(2010-0022369).



\end{document}